\documentclass[fleqn,usenatbib]{mnras}

\usepackage{newtxtext,newtxmath}

\usepackage[T1]{fontenc}
\usepackage{ae,aecompl}

\usepackage{graphicx}	
\usepackage{amsmath}	
\usepackage{amssymb}	
\usepackage[english]{babel}    
\usepackage{subfig}
\usepackage{array, booktabs, makecell}
\usepackage{siunitx}
\usepackage[version=4]{mhchem}
\allowdisplaybreaks

\title[Cold molecular gas around Hot DOGs]{Cold Molecular Gas and Free-Free Emission from Hot, Dust-Obscured Galaxies at $z\rm\sim3$}

\author[J.I. Penney et al.]{
J.I. Penney,$^{1}$\thanks{E-mail: jip3@leicester.ac.uk}
A. W. Blain,$^{1}$
R. J. Assef,$^{2}$
T. Diaz-Santos,$^{2,3,4}$
J. Gonz{\'a}lez-L{\'o}pez,$^{2,5}$
\and C. -W. Tsai,$^{6}$
M. Aravena,$^{2}$ 
P. R. M. Eisenhardt,$^{7}$
S. F. Jones,$^{8}$
H. D. Jun,$^{9}$
\and M. Kim,$^{10}$
D. Stern,$^{7}$
J. Wu$^{6}$
\\
$^{1}$Department of Physics and Astronomy, University of Leicester, University Road, Leicester, LE1 7RH, UK\\
$^{2}$N{\'u}cleo de Astronom{\'i}a de la Facultad de Ingenier{\'i}a y Ciencias, Universidad Diego Portales, Av. Ej{\'e}rcito Libertador 441, Santiago, Chile\\
$^{3}$Chinese Academy of Sciences South America Center for Astronomy (CASSACA), National Astronomical Observatories, CAS,\\ Beijing 100101, China\\
$^{4}$Institute of Astrophysics, Foundation for Research and Technology—Hellas, Heraklion, GR-70013, Greece \\
$^{5}$Las Campanas Observatory, Carnegie Institution of Washington, Casilla 601, La Serena, Chile\\
$^{6}$National Astronomical Observatories, Chinese Academy of Sciences, 20A Datun Road, Chaoyang District, Beijing, 100012, China\\
$^{7}$Jet Propulsion Laboratory, California Institute of Technology, 4800 Oak Grove Drive, CA 91109, USA\\
$^{8}$Department of Space, Earth, and Environment, Chalmers University of Technology, Onsala Space Observatory, SE-43992, Onsala, Sweden\\
$^{9}$School of Physics, Korea Institute for Advanced Study, 85 Hoegiro, Dongdaemun-gu, Seoul 02455, Korea\\
$^{10}$Department of Astronomy and Atmospheric Sciences, Kyungpook National University, Daegu 702-701, Korea
}

\date{Accepted XXX. Received YYY; in original form ZZZ}

\pubyear{2019}

\begin{document}
\label{firstpage}
\pagerange{\pageref{firstpage}--\pageref{lastpage}}
\maketitle

\begin{abstract}
We report on observations of redshifted CO(1--0) line emission and observed-frame $\rm\sim30\,$GHz radio continuum emission from five ultra-luminous, mid-IR selected hot, Dust-Obscured Galaxies (Hot DOGs) at $z\rm\gtrsim3$ using the Karl G. Jansky Very Large Array. We detect CO(1--0) line emission in all five Hot DOGs, with one of them at high signal to noise. We analyse FIR-radio spectral energy distributions, including dust, free-free and synchrotron emission for the galaxies. We find that most of the $\rm115\,$GHz rest-frame continuum is mostly due to synchrotron or free-free emission, with only a potentially small contribution from thermal emission. We see a deficit in the rest-frame $\rm115\,$GHz continuum emission compared to dusty star-forming galaxies (DSFGs) and sub-millimetre galaxies (SMGs) at high redshift, suggesting that Hot DOGs do not have similar cold gas reserves compared with star-forming galaxies. One target, W2305-0039, is detected in the FIRST $\rm1.4\,GHz$ survey, and is likely to possess compact radio jets. We compare to the FIR-radio correlation, and find that at least half of the Hot DOGs in our sample are radio-quiet with respect to normal galaxies. These findings suggest that Hot DOGs have comparably less cold molecular gas than star-forming galaxies at lower, $z\rm\sim2$ redshifts, and are dominated by powerful, yet \emph{radio-quiet} AGN.
\end{abstract}



\section{Introduction}\label{Intro}
Galaxy evolution models currently predict that the formation of most massive galaxies ($\rm M_{*}\gtrsim10^{12}\,M_{\odot}$) involve major mergers at high redshift \citep[see][and references therein]{CJConselice14}. This major-merger theory could be key to understanding how elliptical galaxies form \citep{AToomre77,JEBarnes98,JKormendy12}. Mergers cause large quantities of gas to be forced into the central regions of the galaxies, with some accreting onto a central super-massive black hole (SMBH) \citep{PFHopkins08}. Mergers produce a burst of high-mass star formation as cold gas condenses in giant molecular clouds, with some expelled in winds \citep{DBSanders88a,JEBarnes92,FSchweizer98, FaucherG12}.

\begin{table*}
    \centering
    \caption{Properties of the sample in this paper, showing the CO(1--0) position, redshift, major and minor axis of the synthesized beam in the VLA images, and the position angle of the synthesized beam. All targets are roughly the same size as the beam. Redshifts, discussed in Section~\ref{Results}, are based on optical and near-IR spectroscopy \citep{JWu12,CWTsai15}, although W0126-0529 has an ambiguous redshift detection and the most recent redshift has been placed in parentheses \citep[see][]{HJun20}. Here, ($\rm^{*}$) shows an ALMA redshift identification from CO($J$=6--5) emission.}
    \begin{tabular}{c c c c c c c}
         \hline \\[-2ex]
         WISE Designation & RA (J2000) & DEC (J2000) & $z$ & Major (arcsec) & Minor (arcsec) & PA (deg) \\
         \hline \\[-2ex]
         W0116$-$0505 & 01:16:01.49 & -05:05:05.1 & 3.173 & 4.00 & 2.62 & $\rm-30.96$\\
         W0126$-$0529 & 01:26:11.95 & -05:29:09.1 & 2.937 (0.8301) & 4.17 & 2.44 & $\rm-36.48$\\
         W0410$-$0913 & 04:10:10.62 & -09:13:05.8 & $\rm3.630^{*}$ & 4.08 & 2.92 & $\rm-9.59$\\
         W0831$+$0140 & 08:31:53.25 & 01:40:10.3 & 3.912 & 3.78 & 3.06 & $\rm-2.69$\\
         W1322$-$0328 & 13:22:32.55 & -03:28:42.7 & 3.043 & 3.37 & 2.52 & $\rm6.28$\\
         W2305$-$0039 & 23:05:25.93 & -00:39:25.3 & 3.106 & 3.61 & 2.64 & $\rm-29.03$\\
         \hline
    \end{tabular}
    \label{tab:objects}
\end{table*}

Radio observations are critical to understanding star formation in galaxies at high redshift, providing high resolution images and estimates of the amount and dynamics of cold star-forming gas within the host galaxy. Molecular hydrogen (henceforth $\rm H_{2}$) gas traces regions of future star formation, though with strongly forbidden transitions it can only be observed at high temperatures ($\rm T>100\,$K). The second most abundant molecule in the universe however, $\rm ^{12}CO$, is readily excited and more easily observed. The lowest $\rm^{12}$CO transition,  $\rm^{12}$CO($J$=1--0) (henceforth CO(1--0)), traces the coldest molecular gas \citep{AOmont07,RJIvison11,ADBolatto13}. CO(1--0) transitions are fainter than the higher $J$-transitions \citep{ADBolatto13}, but should more accurately measure the molecular gas content in galaxies. By assuming $\rm ^{12}CO$ and $\rm H_{2}$ are linked \citep{PMSolomon97,LJTacconi08,ADBolatto13}, the mass of the $\rm H_{2}$ regions can be estimated. Further, detection of the lowest CO(1--0) transitions can be used in tandem with higher transitions (e.g. $J$=4--3) to understand properties such as local density of gas, the star formation rate and temperature of the star-forming gas \citep[e.g.][]{CHPenaloza17}. Thus, from CO(1--0) observations, we can understand the molecular gas content necessary to form stars in high redshift galaxies, potentially isolating star formation in the surrounding galaxy from high-redshift, powerful AGN.

Hot, Dust-Obscured Galaxies \citep[Hot DOGs;][]{JWu12} are hyper-luminous infra-red (IR) galaxies (HyLIRGs\footnote{Here, LIRGs, ULIRGs and HyLIRGs are characterized by a total IR luminosities of $10^{11}$ $<$$L_{\rm8-1000\,\mu m}/L_{\odot}$ $<$$10^{12}$, $10^{12}$ $<$$L_{\rm8-1000\,\mu m}/L_{\odot}$ $<$$10^{13}$ and $\rm L_{8-1000\,\mu m}/L_{\odot}$ $>$$10^{13}$, respectively \citep{DBSanders96}.}) initially discovered using the $\textit{Wide-field IR Survey Explorer}$ \citep[$\textit{WISE}$;][]{ELWright10} All-Sky Survey \citep{RMCutri12} by selecting for bright detections in the W3 ($\rm12\,\mu$m) and W4 ($\rm22\,\mu$m) bands and faint or no detections in the W1 ($\rm3.4\,\mu$m) and W2 ($\rm4.6\,\mu$m) bands \citep{PRMEisenhardt12}. Follow-up observations by \cite{JWu12} to determine more complete spectral energy distributions (SEDs) suggest significantly higher fractions of hot dust ($\rm\gtrsim60\,$K) than most dusty galaxies, with a peak in their SEDs at rest-frame $\rm\sim22\,\mu$m, and high IR luminosities \citep[$\rm>10^{13}\,L_{\odot}$;][]{SFJones14,CWTsai15}. 

Optical spectra typically reveal narrow AGN lines, explaining their high luminosities and dust temperatures \citep{CWTsai15}. Follow-up observations using the $\textit{Spitzer Space Telescope}$ \citep[henceforth $\textit{Spitzer}$;][]{MWWerner04} show mid-IR SEDs consistent with obscured AGN emission \citep{RJAssef15}. $\rm850\,\mu$m SCUBA-2 observations \citep{SFJones14} show these galaxies inhabit overdense environments of sub-millimeter galaxies \citep[SMGs;][]{AWBlain02}, with number densities within $\rm1.5\,'$ of the galaxies $\rm\sim3$ times higher than blank fields. \cite{RJAssef15} used warm-$\textit{Spitzer}$ observations to show these galaxies reside in regions with elevated surface densities of IRAC colour-selected galaxies, consistent with the rich environments of radio-loud AGN. For instance, the Clusters around Radio-Loud AGN \citep[CARLA;][]{DW13} survey found that $\rm92\%$ of the $\rm\sim400$ radio-loud AGN they observed with $\textit{Spitzer}$ are overdense compared to the field, with most ($\rm55\%$) overdense at the $\rm\geq2\sigma$ level on $\rm1\,'$ radius scales. Given the similar overdensity for Hot DOGs compared with radio-loud galaxies in \cite{DW13}, it is of interest to understand the $\rm\sim1.4\,GHz$ radio properties of Hot DOGs and whether they are similar to radio-loud galaxies or to the overall AGN population.

The Karl G. Jansky Very Large Array \citep[VLA;][]{EVLA} has full coverage between 1 and $\rm50\,$GHz with its upgrade in 2011, and it can now make unprecedented radio observations over a large frequency range. Here, we present CO(1--0) and rest-frame $\rm115.3\,$GHz continuum observations of a sample of six hyper-luminous Hot DOGs from \cite{CWTsai15} using the lowest resolution and most sensitive configuration of the VLA (Table~\ref{tab:objects}). These galaxies were selected for their high luminosities ($\rm L_{bol}\geq 10^{14}\,L_{\odot}$). We use the VLA to obtain CO(1--0) luminosities and rest frame $\rm115.3\,$GHz continuum fluxes of these galaxies to infer $\rm H_{2}$ masses, as well as free-free and thermal activity of these highly obscured galaxies at the peak 
redshift of major mergers, which is before the peak redshift for starbursts and AGN activity \citep{PFHopkins08}. 

Section~\ref{Observations} describes details of the observations, Section~\ref{Results} discusses the main results of these observations and Section~\ref{SEDs} uses archival data in conjunction with new radio observations to construct SEDs of the galaxies. Section~\ref{Discussion} discusses these findings. Throughout, we assume a cosmology of $\rm H_{0}=70\,km\,s^{-1}\,Mpc^{-1}$, $\rm\Omega_{m}=0.30$ and $\rm\Omega_{\Lambda}=0.70$.

\begin{figure*}
\captionsetup[subfigure]{labelformat=empty}
\begin{tabular}{cc}
\subfloat[][]{\includegraphics[trim={0.0cm 0.0cm 1.0cm 0.2cm}, clip, height=0.3\textwidth, width=0.32\textwidth] {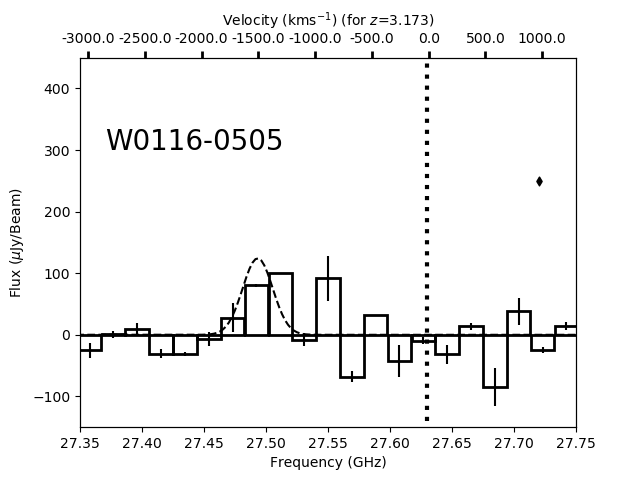}\label{W0116_Spectrum}}\hspace{0.05cm}
\subfloat[][]{\includegraphics[trim={0.0cm 0.0cm 1.0cm 0.2cm}, clip, height=0.3\textwidth, width=0.32\textwidth] {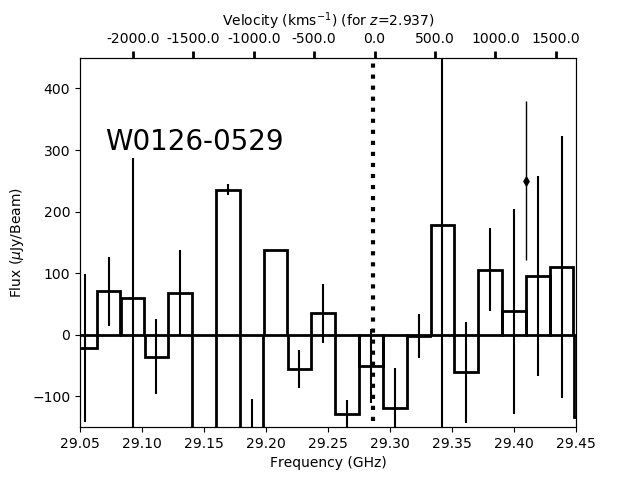}\label{W0126_Spectrum}}\hspace{0.05cm}
\subfloat[][]{\includegraphics[trim={0.0cm 0.0cm 1.0cm 0.2cm}, clip, height=0.3\textwidth, width=0.32\textwidth] {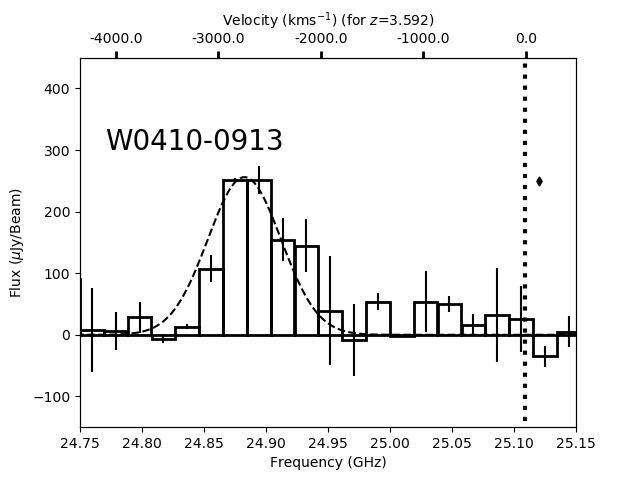}\label{W0410_Spectrum}}\\[-6ex]
\subfloat[][]{\includegraphics[trim={0.0cm 0.0cm 1.0cm 0.2cm}, clip, height=0.3\textwidth, width=0.32\textwidth] {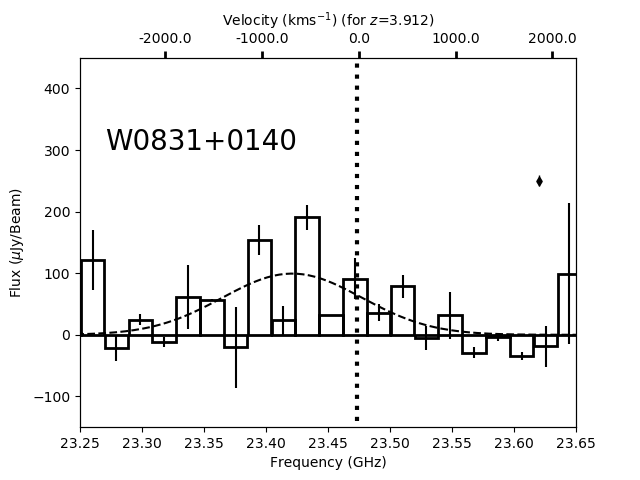}\label{W0831_Spectrum}}\hspace{0.05cm}
\subfloat[][]{\includegraphics[trim={0.0cm 0.0cm 1.0cm 0.2cm}, clip, height=0.3\textwidth, width=0.32\textwidth] {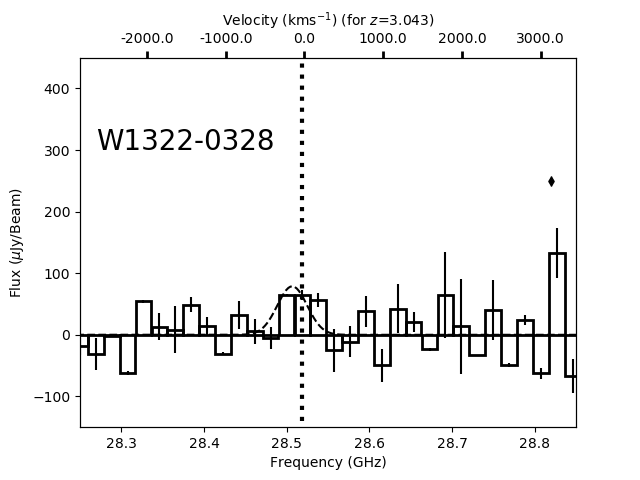}\label{W1322_Spectrum}}\hspace{0.05cm}
\subfloat[][]{\includegraphics[trim={0.0cm 0.0cm 1.0cm 0.2cm}, clip, height=0.3\textwidth, width=0.32\textwidth] {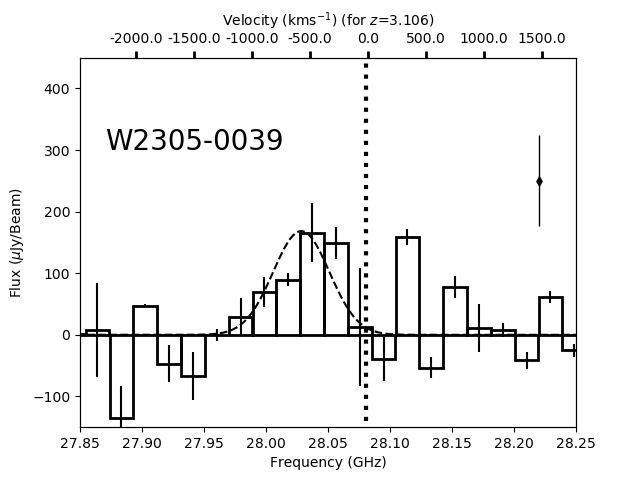}\label{W2305_Spectrum}}\\[-6ex]
\end{tabular}
\caption{Continuum subtracted spectra of the six sources in our sample. A Gaussian has been added to illustrate the CO(1--0) line. An additional error bar in the upper right-hand corner has been included to show the average continuum level for each field. Dotted vertical lines indicate the expected CO(1--0) wavelength based on optical/IR redshifts \citep[see][]{CWTsai15}. Frequency ranges have been chosen to highlight the line emission of the galaxies.}
\label{Spectra_Images}
\end{figure*}

\section{Observations and Data Reduction}\label{Observations}
Observations of six sources (Table~\ref{tab:objects}) were carried out in $\rm1-2\,$hr blocks during generally excellent weather conditions between February and April 2017. The most compact baseline, D configuration, was used for the observations due to its greater sensitivity. Two observation blocks were used for each galaxy in our sample, except for W0126-0529, which had a single $\rm\sim2\,$hr observation. The K-band ($\rm\nu$=18.0--26.5$\,$GHz) receiver was used for W0410-0913 and W0831+0140, and Ka-band ($\rm\nu$=26.5--40.0$\,$GHz) for W0116-0505, W0126-0529, W1322-0328 and W2305-0039. A single polarization was used, and the flux calibrators were 3C48 (for W0116-0505, W0126-0529 and W2305-0039), 3C138 (for W0410-0913 and W0831+0140), and 3C286 (for W1322-0328). The observations produced $\rm\sim1\,$GHz bandwidth spectra centred on the expected CO(1--0) emission frequency \citep[$\rm\nu_{rest}=115.3\,$GHz;][]{CDMorton94} based on redshifts from optical/IR spectra presented in \cite{JWu12} and \cite{CWTsai15}.

Calibration and imaging were carried out using the Common Astronomy Software Application \citep[CASA;][]{CASA}\footnote{https://casa.nrao.edu}. Calibrations were made using the pipeline\footnote{https://science.nrao.edu/facilities/vla/data-processing/pipeline/scripted-pipeline} for all fields except W1322-0328, in which a custom calibration method detailed in the online calibration cookbook\footnote{https://casaguides.nrao.edu/index.php/\newline TWHydraBand7$\_$Imaging$\_$4.3} was used for one of the observation blocks because of artefacts arising from a bad reference antenna using the pipeline. No additional calibrations were required following the pipeline calibrations.

The six fields were cleaned using CASA's \textsc{tclean} package. Initial dirty spectral maps were made to estimate the positions of the galaxies and the frequency of emission. Continuum subtraction was then carried out using the \textsc{uvcontsub} package. Cleaning and imaging of all fields was made using \textsc{hogbom} deconvolution, with \textsc{briggs} weighting and $\rm\textsc{robust}=2.0$ to produce a natural weighting. All spectral cubes were constructed using $\rm6\,$MHz ($\rm50\,km\,s^{-1}$) resolution to boost the signal-to-noise ratio (SNR). Spectra were extracted using a single $\rm0.05\,''$ pixel centred on the positions of the peak CO(1--0) emission, determined using CASA's \textsc{imfit} routine, as shown in Table~\ref{tab:objects}.

\section{Observational Results}\label{Results}
The spectra for the six targets in our sample are shown in Fig.~\ref{Spectra_Images}. Errors are calculated from each individual channel of the spectra, and is $\rm\sim10-20\%$ of the peak flux for each frame. W0410-0913 has the strongest emission line, with a SNR of 4.21 at the peak and redshift of $z\rm=3.633\pm0.005$, \citep[compared with an optical redshift of $z\rm=3.592\pm0.002$ from][]{JWu12}. The CO(1--0) redshift matches the redshift from the CO($J$=6--5) line detection using ALMA (Gonz{\'a}lez-L{\'o}pez et al. in prep.). This suggests that the gas associated with the lower, optical redshift could be emitted from outflows at $\rm\sim3000\,km\,s^{-1}$. This is expected, given work by \cite{TDSantos18}, which saw blue-shifted lines from $\rm Ly\alpha$ in another Hot DOG, W2246-0526. 

\begin{figure*}
    \centering
    \includegraphics[trim={0.0cm 0.0cm 1.5cm 0.0cm}, clip, width=\textwidth] {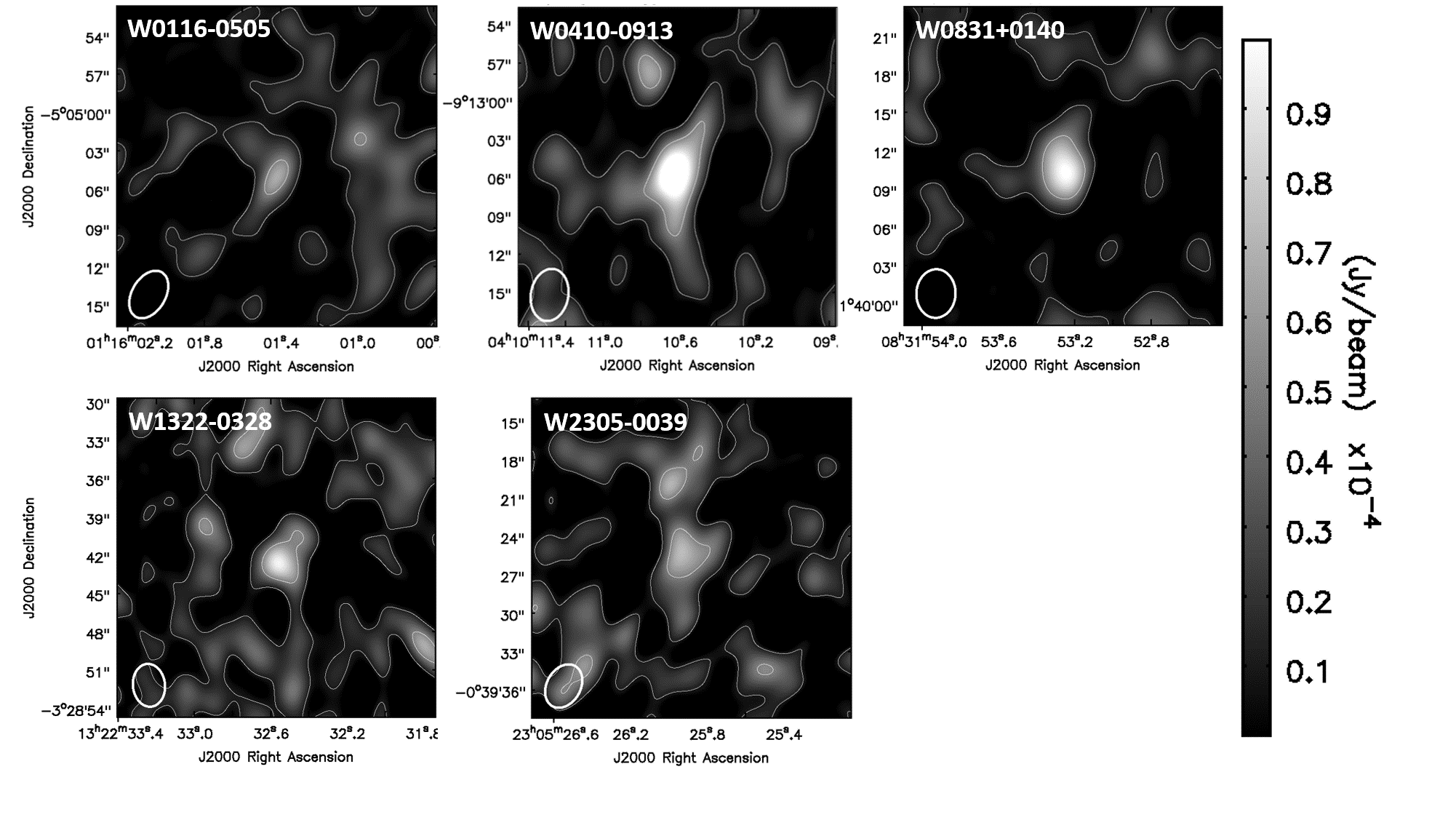}\\[-6ex]
    \caption{Emission centred images (moment=0) for W0116-0505, W0410-0913, W0831+0140, W1322-0328 and W2305-0039 with continuum subtraction. Contours show the $\rm0.1\,\mu$Jy, $\rm0.5\,\mu$Jy and $\rm1\mu$Jy levels. An ellipse in the bottom left-hand corner of each image illustrates the synthesized beam for each image.}
    \label{Emission_images}
\end{figure*}

The other galaxies in our sample show evidence of weak CO(1--0) emission, with the exception of W0126-0529, which has the least confident redshift identification in \cite{CWTsai15}, based on a weak emission line assumed to be Ly$\alpha$. \cite{HJun20} determined a lower redshift for this galaxy using X-SHOOTER data: $z\rm=0.8301$. From our VLA observations, we confirm that there is no CO(1--0) emission at a redshift of $z\rm=2.937$. Based on these findings, the galaxy does not meet the luminosity selection criterion for the sample, and has been omitted from the rest of the discussion of the paper. All of the results for this object have been included in Appendix~\ref{Appendix}. The emission line from W0116-0505 is at a lower frequency than expected, by $\rm\delta\nu$=$\rm0.09\,$GHz, a redshift of $z\rm=3.187\pm0.005$, compared to $z$=$\rm3.173\pm0.002$ in \cite{JWu12}. The velocity difference of $\rm\sim1400\,km\,s^{-1}$ suggests that the optically detected gas in W0116-0505 is also associated with out-flowing material. The CO(1--0) redshifts for W0831+0140, W1322-0328 and W2305-0039 agree with the optical/IR redshifts in \cite{CWTsai15}, within $\rm\delta v \sim500\,km\,s^{-1}$. The full width at half-maximum (FWHM) of the fitted Gaussian for each of the galaxies is between $\rm\sim140-470\,km\,s^{-1}$ (see Table~\ref{tab:Luminosity1}), much broader than the Milky Way CO(1--0) emission \citep[$\rm\lesssim20\,km\,s^{-1}$;][]{KSheth08}.

The images of the CO(1--0) emission are shown in Fig.~\ref{Emission_images} in the spectral window (henceforth $spw$) centred on the detected emission line and the entire $spw$ centred on the emission line is used. All galaxies show line and continuum emission centred on the target. All fields have a $\rm SNR\gtrsim3$ (mean $\rm SNR=4.49$), showing that the positions of the emission lines are similar to the expected positions from previous observations of these galaxies, implying they are unlikely to be significantly affected by background noise. Despite these relatively low SNRs, we can assume the VLA emission is attributed to these galaxies given the prior that these galaxies are detected at these positions at other wavelengths.

Using CASA's \textsc{immoments}, we imaged the velocity fields of the five Hot DOGs (moments$\rm=1$) to investigate any motion of molecular gas using the $spw$ associated with the emission lines. We find no signs of rotational motion of the gas, as expected given the modest resolution and SNR of our observations. Unsurprisingly, there is no evidence for large-scale bulk motion of cold molecular gas around these galaxies on $\rm20-50\,kpc$ ($\rm3-6\,''$) scales. The cold molecular gas could have a velocity gradient in the Hot DOGs, as shown by the width of the CO(1--0) line in Fig.~\ref{Spectra_Images}, suggesting that the cold molecular gas has not been evaporated by the emission of the central AGN. However, we are unable to detect the motion of the cold molecular gas given the current data. Greater spatial resolution and sensitivity are needed to understand whether there is any outflow of cold molecular CO gas responsible for the change in expected emission from W0116-0505.

\begin{figure*}
    \centering
    \includegraphics[trim={0.0cm 0.0cm 0.0cm 0.0cm}, clip, width=\textwidth] {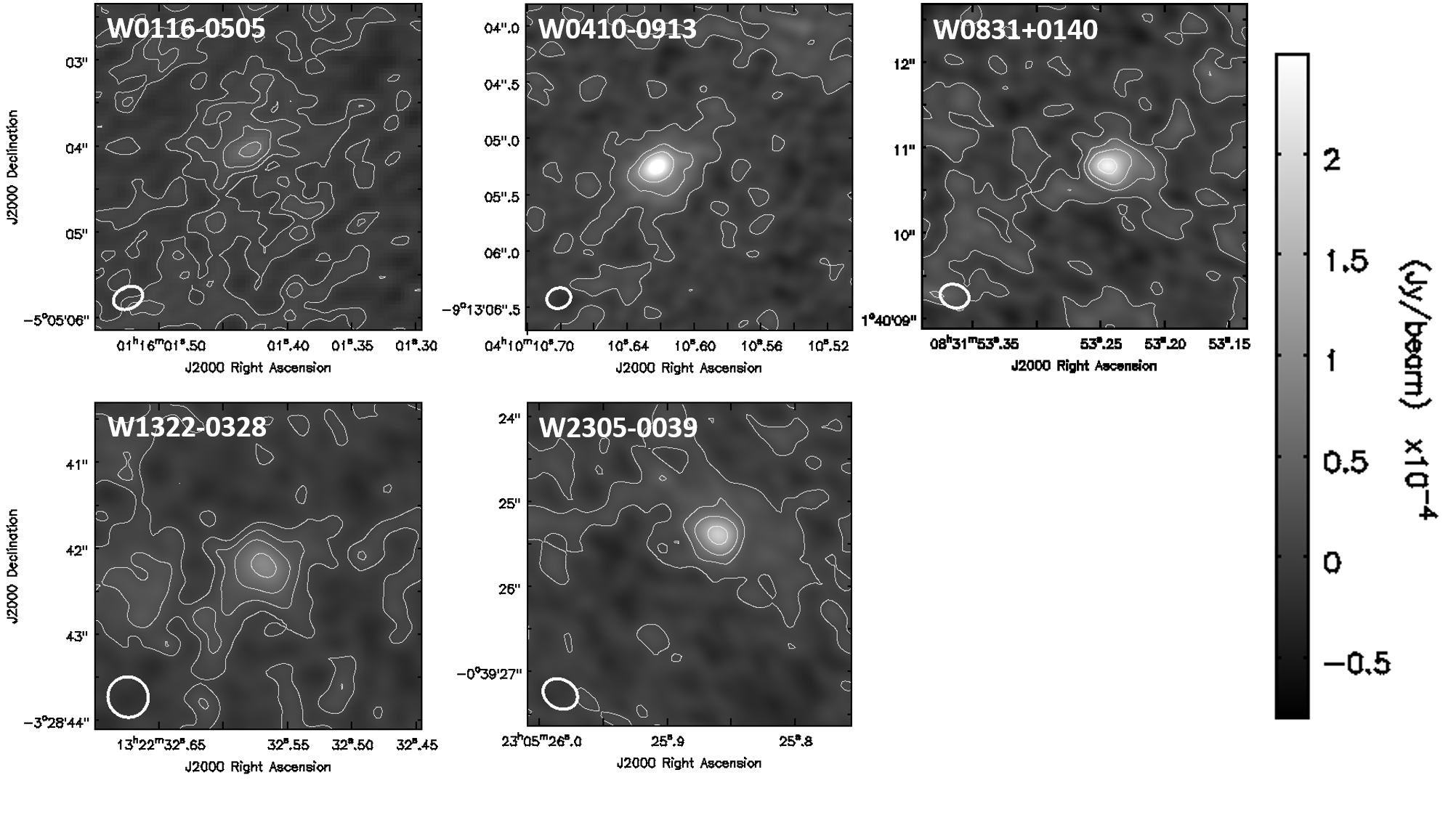}\\[-5ex]
    \caption{Continuum images of the five Hot DOGs in the sample observed using ALMA in Band-3 ($\rm2.6-3.6\,mm$) and Band-5 ($\rm1.4-1.8\,mm$) for W0410-0913. Contours show the $\rm0.1\,\mu$Jy, $\rm0.5\,\mu$Jy, $\rm1\,\mu$Jy and $\rm2\,\mu$Jy levels. An ellipse in the bottom left-hand corner of each image illustrates the synthesized beam for each image. A colour-bar has been provided to show the flux levels across the images.}
    \label{ALMA_Continuum}
\end{figure*}

We now compare our VLA observations with independent observations using ALMA (Gonz{\'a}lez-L{\'o}pez et al. in prep.). These ALMA observations primarily looked at potential companion sources using higher $J$-components, looking at the warmer molecular gas in the environments of Hot DOGs. ALMA observations of W0116-0505, W0831+0140, W1322-0328 and W2305-0039 were taken as part of project 2017.1.00358.S, using ALMA's Band-3 ($\rm2.6-3.6\,mm$) to image CO($J$=4--3) emission. Additional observations were made for W0410-0913 using Band-5 ($\rm1.4-1.8\,mm$) to image CO($J$=5--4) emission as part of project 2017.1.00908.S. The data were processed with the ALMA calibration pipeline in CASA \citep{JPMcMullin07}, using scripts provided by ALMA. The images and data cubes were created using the task \textsc{tclean} with natural weighting. The continuum image was obtained using the multi-frequency synthesis (mfs) mode with all the line-free $spw$. Continuum emission was then subtracted from the visibilities using \textsc{uvcontsub}. All of the continuum images and data cubes were cleaned down to the $\rm2\sigma$ level using auto-masking. The typical RMS value for the continuum maps was $\rm9-15\,\mu Jy\,beam^{-1}$ with a beam size ranging from $\rm0.2-0.5\,''$. Continuum measurements were carried out using the task \textsc{imfit}, and were checked using aperture photometry, which uncovered similar results. Fig.~\ref{ALMA_Continuum} shows the continuum maps for the ALMA observations.

From the VLA observations in Fig.~\ref{Emission_images}, we see a potential companion to W2305-0039 within $\rm\sim5\,''$ of the Hot DOG (RA=23:05:26.00, DEC=--00:39:19.91) and is of interest due to the proximity to the target. Inspecting the VLA spectrum at the position of this additional source, we find a peak at a frequency corresponding to the same redshift of CO(1--0) emission as W2305-0039 ($\rm L'_{CO(1-0)}=0.88\pm0.50\times10^{10}\,K\,km\,s^{-1}\,pc^{2}$). Comparing with ALMA observations of warmer CO($J$=4--3) gas from Gonz{\'a}lez-L{\'o}pez et al. (in prep.) at this position we find a lack of peak in the spectrum, suggesting that the companion is not composed of warmer CO gas. The W2305-0039 companion is likely to either be a site of additional cold molecular gas within the Hot DOG, or associated with the synchrotron emission seen in the FIRST\footnote{Faint images of the radio sky at twenty-cm} survey (see Section~\ref{Radio_Emission_Hot_DOGs}). We discuss the luminosity of these Hot DOGs with respect to observations of other galaxies in Section~\ref{CO_Lum}.

\begin{figure}
    \centering
    \includegraphics[trim={0.2cm 0.0cm 1.0cm 1.0cm}, clip, width=0.47\textwidth]{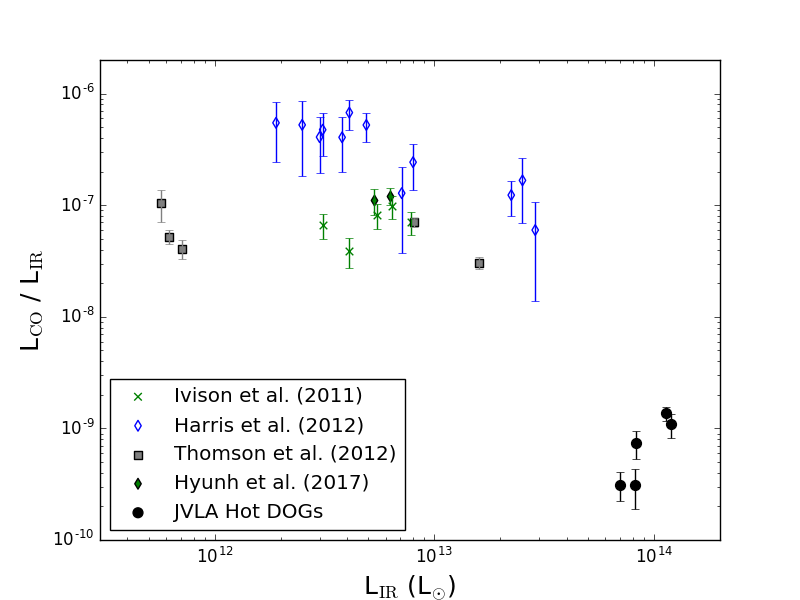}\label{Lco_Lir_Lir_fig}\\
    \includegraphics[trim={0.2cm 0.0cm 1.0cm 1.0cm}, clip, width=0.47\textwidth]{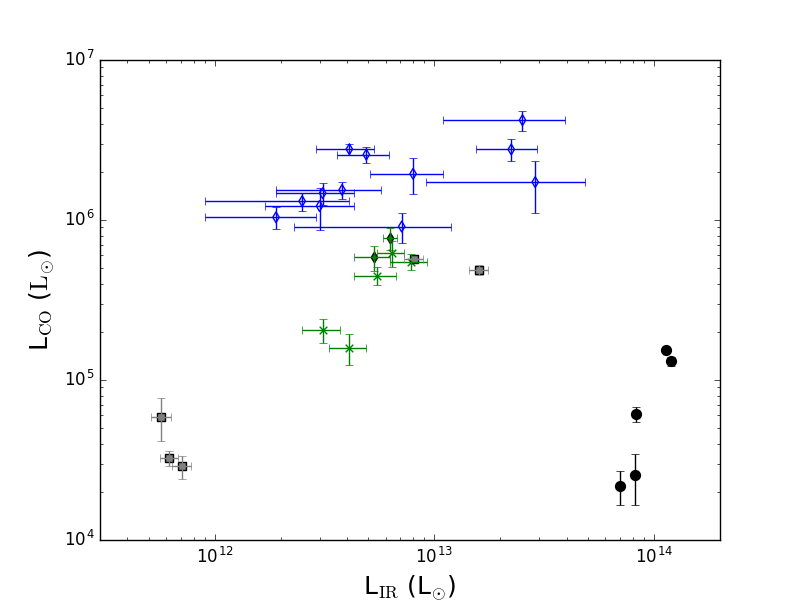}\label{LcodL_LirdL_fig}\\
    \includegraphics[trim={0.2cm 0.0cm 1.0cm 1.0cm}, clip, width=0.47\textwidth]{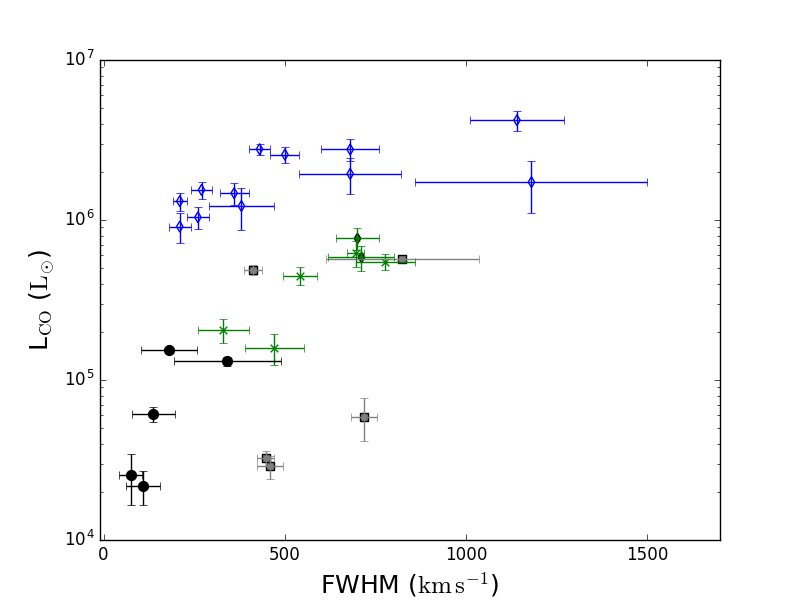}\label{Lco_FWHM_fig}\\[-1ex]
    \caption{Comparison of the $\rm L_{CO(1-0)}$ properties of the galaxies in our sample with other galaxies at high redshift. Top: The $\rm L_{CO(1-0)}$/$\rm L_{IR}$ ratio against $\rm L_{IR}$. Middle: Comparison of the $\rm L_{CO(1-0)}$ against $\rm L_{IR}$. Bottom: $\rm L_{CO(1-0)}$ against the FWHM of the CO(1--0) emission line. In each, we compare against $z\rm=2.2-2.5$ ALMA-selected SMGs \citep{RJIvison11} (green crosses), lens-corrected DSFGs (2.1$\leq z\leq$3.5) \citep{AIHarris12} (blue empty diamonds), strongly lensed (corrected) SMGs at $z\rm=2.5-3$ \citep{APThomson12} (grey squares), and ALMA selected, Australian Telescope Compact Array observed galaxies at $z\rm=2.0232$ and $\rm2.1230$ respectively \citep{MTHyunh17} (green filled diamonds). VLA-observed Hot DOGs are illustrated by black filled circles.}
    \label{L_co_properties_fig}
\end{figure}

\begin{table*}
    \centering
    \caption{Bolometric luminosity \protect\citep{CWTsai15}, CO(1--0) luminosity, FWHM of the Gaussian fit for CO(1--0) and estimated $\rm H_{2}$ masses. The $\rm L_{bol}$ values are likely to be lower limits, and the uncertainties of $\rm14\%-26\%$ are in consideration of the difference between the power-law connection method \citep{CWTsai15} and a more sophisticated continuous temperature model. See Section~\ref{CO_Lum} for values marked with $^{\alpha}$ for calculation.}
    \begin{tabular}{c c c c c}
         \hline \\[-2ex]
         ID & $\rm L_{bol}$ $\rm(10^{13} L_{\odot})$ & $\rm L'_{CO(1-0)}$ $\rm(10^{10} K\,km\,s^{-1}\,pc^{2})$ $^{\alpha}$ & FWHM$\rm_{CO(1-0)}$ ($\rm km\,s^{-1}$) & $\rm M(H_{2})$ ($\rm10^{10} M_{\odot}$) $^{\alpha}$ \\
         \hline \\[-2ex]
         W0116$-$0505 & 11.7 $\pm$ 2.0 & 0.43 $\pm$ 0.15 & 75 $\pm$ 32 & 0.34 $\pm$ 0.12 \\
         W0410$-$0913 & 16.8 $\pm$ 2.4 & 2.60 $\pm$ 0.06 & 180 $\pm$ 77 & 2.08 $\pm$ 0.05 \\
         W0831$+$0140 & 18.0 $\pm$ 4.1 & 2.21 $\pm$ 0.15 & 342 $\pm$ 146 & 1.77 $\pm$ 0.12 \\
         W1322$-$0328 & 10.1 $\pm$ 1.4 & 0.37 $\pm$ 0.09 & 107 $\pm$ 47 & 0.30 $\pm$ 0.07 \\
         W2305$-$0039 & 13.9 $\pm$ 3.6 & 1.04 $\pm$ 0.11 & 137 $\pm$ 59 & 0.83 $\pm$ 0.09 \\
         \hline
    \end{tabular}
    \label{tab:Luminosity1}
\end{table*}

\begin{table*}
    \centering
    \caption{Continuum values for the five Hot DOGs observed with ALMA ($\rm S_{ALMA}$) (see Fig.~\ref{ALMA_Continuum}) and the continuum values for all five Hot DOGs for the VLA ($\rm S_{1\,cm}$). $\rm\nu_{VLA}$ shows the range of frequencies observed in the VLA observations. The final column gives the flux ratio of the line to continuum for each Hot DOG.}
    \begin{tabular}{c c c c c}
        \hline\\[-2ex]
         ID & $\rm S_{ALMA}$ ($\mu$Jy) & $\rm S_{1\,cm}$ ($\rm\mu$Jy) & $\rm\nu_{VLA}$ (GHz) & $\rm S_{CO(1-0)}$/$\rm S_{1\,cm}$ \\
         \hline \\[-2ex]
         W0116$-$0505 & 241 $\pm$ 49 & 5.36 $\pm$ 2.07 & 27.6--28.6 & $\rm29.0\pm15.1$ \\
         W0410$-$0913 & 716 $\pm$ 66 & 6.01 $\pm$ 1.21 & 24.5--25.5 & $\rm131.9\pm26.7$ \\
         W0831$+$0140 & 435 $\pm$ 47 & 8.67 $\pm$ 1.76 & 23.0--24.0 & $\rm66.7\pm14.3$ \\
         W1322$-$0328 & 140 $\pm$ 26 & 15.5 $\pm$ 0.84 & 27.9--28.9 & $\rm3.8\pm0.9$ \\
         W2305$-$0039 & 283 $\pm$ 35 & 73.8 $\pm$ 3.55 & 27.5--28.5 & $\rm7.0\pm0.8$ \\
         \hline
    \end{tabular}
    \label{tab:Luminosity2}
\end{table*}

\subsection{CO(1--0) Luminosity}\label{CO_Lum}
Using the emission spectra in Fig.~\ref{Spectra_Images}, we determine the CO(1--0) luminosities of the lines. Using the same method as \cite{PMSolomon97}, we derive the luminosity from the integrated flux of the emission line using Eqn.~\ref{equation:luminosity}:

\begin{equation}\label{equation:luminosity}
    \rm L'_{CO(1-0)} = 3.25\times10^{7} \times S_{CO}\delta V \times\frac{D_{L}^{2}}{\nu_{obs}^{2}(1+\textit{z})^3} ,
\end{equation}

\noindent where $S_{CO}$ is the integrated flux density of the line (Jy), $\delta V$ is the velocity width ($\rm km\,s^{-1}$) and $D_{L}$ is the luminosity distance to the galaxy (Mpc). $\rm L'_{CO(1-0)}$ is in units of $\rm K\,km\,s^{-1}\,pc^{2}$, which can be converted to $\rm L_{CO(1-0)}$ (in units of $\rm L_{\odot}$) by $\rm L/L'=(8\pi k_{B}/c^{2})\nu_{rest}^{3}$, where $k_{B}$ is the Boltzmann constant and $c$ is the speed of light. Assuming that the CO gas and $\rm H_{2}$ gas originate from the same reservoir \citep{RJIvison10}, such that the CO gas trace the $\rm H_{2}$ gas, we can convert the luminosity of the CO(1--0) gas to the $\rm H_{2}$ mass using:

\begin{equation}
    \rm M(H_{2}) = \alpha\, L'_{CO} ,
\end{equation}

\begin{figure*}
    \centering
    \includegraphics[trim={0.0cm 0.0cm 0.0cm 0.0cm}, clip, width=\textwidth] {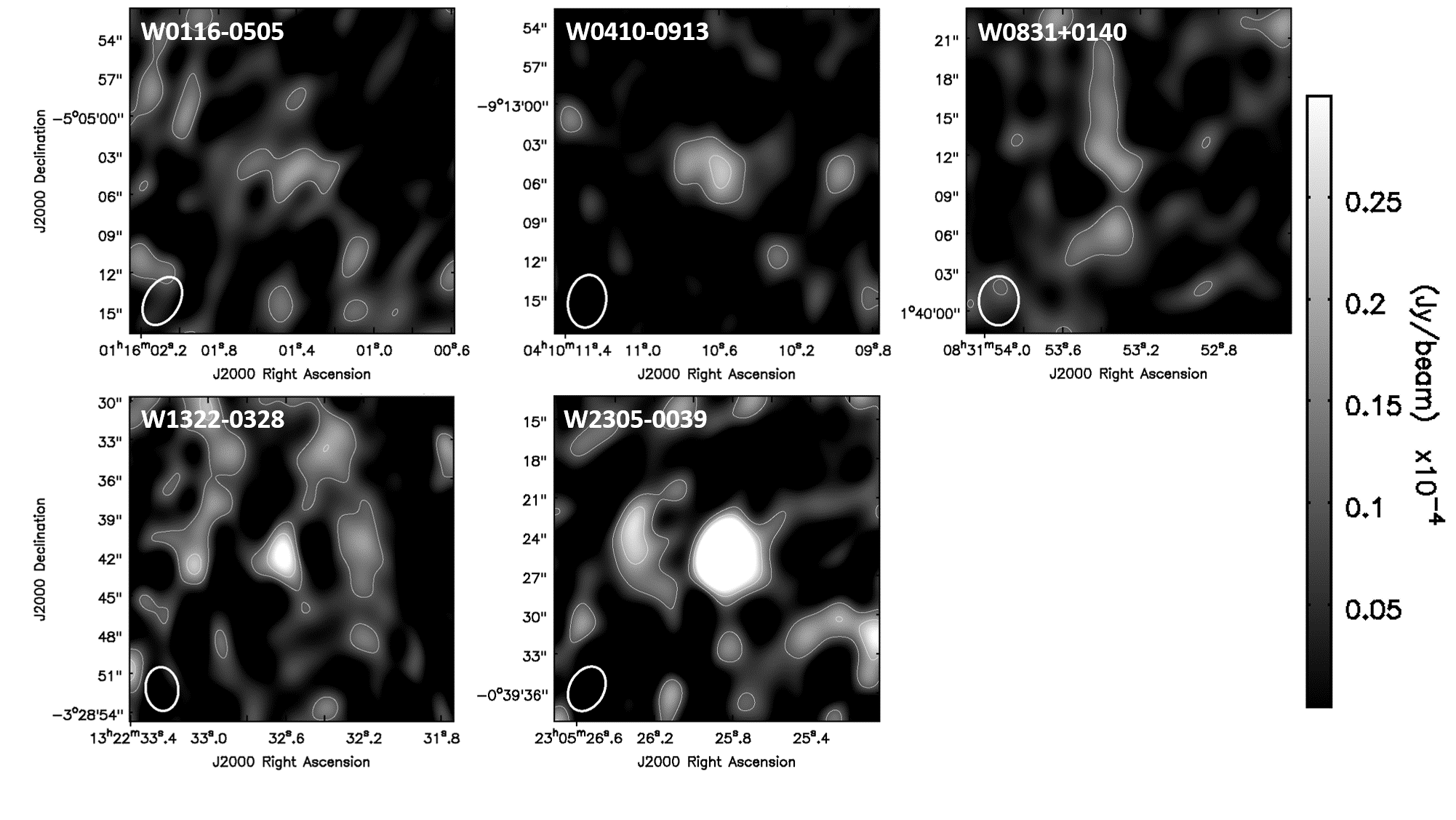}\\[-6ex]
    \caption{Continuum images (moment=0) for (left to right, top to bottom) W0116-0505, W0410-0913, W0831+0140, W1322-0328 and W2305-0039. Contours show the $\rm0.1\,\mu$Jy, $\rm0.2\,\mu$Jy and $\rm0.3\mu$Jy levels, the peak continuum emission for these Hot DOGs. An ellipse in the bottom left-hand corner of each image illustrates the synthesized beam for each image.}
    \label{Continuum_images}
\end{figure*}

\noindent where $\rm M(H_{2})$ is the mass of the $\rm H_{2}$ region in $\rm M_{\odot}$, $L'_{CO}$ is the luminosity of the emission line in $\rm K\,km\,s^{-1}\,pc^{2}$ and $\rm\alpha\sim0.8$ is the conversion factor, valid for ULIRGs and mergers \citep{DDownes98,ADBolatto13}. The results are shown in Table~\ref{tab:Luminosity1} for the five Hot DOGs in our sample. We see CO(1--0) luminosities $\rm\gtrsim0.4\times10^{10}\,K\,km\,s^{-1}\,pc^{2}$, similar to findings by \cite{RJIvison10} and \cite{DARiechers11} for ULIRGs at similar redshifts. We compare the properties of the CO(1--0) luminosity with galaxies at similar redshifts in Fig.~\ref{L_co_properties_fig}. We see that the Hot DOGs in our sample do not follow the expected trend with respect to the $\rm\frac{ L_{CO(1-0)}}{L_{IR}}$ ratio, suggesting that these galaxies are significantly more luminous in the IR than expected based on their $\rm L_{CO(1-0)}$. This implies that the cold gas reservoirs in these Hot DOGs are different to other galaxy populations considered here. However, it should be noted that the $\rm L_{IR}$ is boosted above what would be expected for galaxies, supported by Fig.~\ref{L_co_properties_fig}. We find a mean $\rm\frac{L_{CO(1-0)}}{L_{IR}}$ value of $\rm7.6\pm1.7\times10^{-10}$ which is significantly lower than the value of $\rm4.2\pm0.5\times10^{-8}$ found by \cite{APThomson12} for their two SMGs. The $\rm L_{IR}$ error for the 5 Hot DOGs in this sample is estimated by comparing the $\rm L_{IR}$ in \cite{CWTsai15} and a continuous temperature model, and the $\rm L_{bol}$ values shown in Table~\ref{tab:Luminosity1} are likely to be lower. This suggests that the Hot DOGs are significantly more luminous in the IR than would be expected for galaxies at this redshift for a given $\rm L_{CO(1-0)}$.

We see agreement between the Hot DOGs and comparison samples with respect to the width of the CO(1--0) emission, suggesting that the width of the emission does not substantially deviate from expectations with respect to the luminosity of the line. As expected, the $\rm L'_{CO(1-0)}$ emission from these Hot DOGs is fainter than SMGs at $z$=2.2--2.5 in \cite{RJIvison11}, suggesting that the cold molecular gas in the Hot DOGs is less luminous than SMGs at lower redshifts. In Fig.~\ref{Spectra_Images} we see that most of the Hot DOGs in our sample have CO(1--0) line widths $\rm\lesssim500\,km\,s^{-1}$. The full width at zero intensity (FWZI) values reported in \cite{APThomson12}, $\rm\sim500-1000\,km\,s^{-1}$, are significantly greater, showing that the Hot DOG CO(1--0) lines are narrower than in typical SMGs. Comparing to VLA observations of lensed galaxies by \cite{DARiechers11}, which report line widths of $\rm\gtrsim1300\,km\,s^{-1}$, the Hot DOGs have similar line widths to optically selected QSOs. It should be noted that \cite{DFarrah17} show that Hot DOGs are unlikely to show any lensing, and it is therefore unlikely that the deviation in the $\rm L'_{CO}$-$\rm L_{IR}$ relationship shown in Fig.~\ref{L_co_properties_fig} is due to lensing and is most likely attributed to the hot dust emission from an AGN present in this sample.

Follow-up observations by \cite{JWu14} attempted to determine the cold molecular gas content of two Hot DOGs using the sub-millimetre array (SMA) and the combined array for research in millimetre-wave astronomy (CARMA), using CO(3--2) and CO(4--3) transitions. \cite{JWu14} found masses of $\rm<3.3$ and $\rm<2.3\times10^{10}\,M_{\odot}$ for W0149+2350 and W1814+3412 respectively. Comparing these results with the CO(1--0) masses in Table~\ref{tab:Luminosity1}, we see general agreement for W0116-0505, although the other galaxies in our sample are significantly more massive, suggesting Hot DOGs do not have similar cold molecular gas contents. ALMA observations by \cite{LFan18} for three Hot DOGs using CO(4--3) emission (including W0410-0913) suggest $\rm L'_{CO(4-3)}$ in the range of $\rm10^{10}-10^{11}\,L_{\odot}$. Unfortunately, the CO(4--3) line for W0410-0913 in \cite{LFan18} appears to be clipped, such that the full CO(4--3) emission line is outside the observed range, and so we are unable to comment on whether the use of a conversion factor would show agreement with this work. However, we see that the CO(4--3) line for W0410-0913 appears in the same velocity range as expected from these observations, but has a significantly broader FWHM in the CO(4--3) observations ($\rm330-560\,km\,s^{-1}$) compared to the CO(1--0) observations in this work ($\rm180\,km\,s^{-1}$).

\subsection{Rest Frame $\rm115.3\,$GHz Continuum}
We measure the continuum flux for the five Hot DOGs in our sample using the $spw$s away from the emission line, $\rm\sim0.83\,GHz$ in bandwidth and use the same single pixel position used to extract the spectra for each target. The images of the continuum are shown in Fig.~\ref{Continuum_images} and continuum fluxes are listed in Table~\ref{tab:Luminosity2}. W2305-0039 has the brightest continuum emission, with the other four more weakly detected. The significantly greater continuum value for W2305-0039 could suggest additional dust or synchrotron/free-free emission processes, as discussed in Section~\ref{Radio_Emission_Hot_DOGs}. 

\begin{figure*}
\captionsetup[subfigure]{labelformat=empty}
\subfloat[][]{\includegraphics[trim={0.4cm 0.05cm 1.5cm 0.2cm}, clip, height=0.35\textwidth, width=0.49\textwidth] {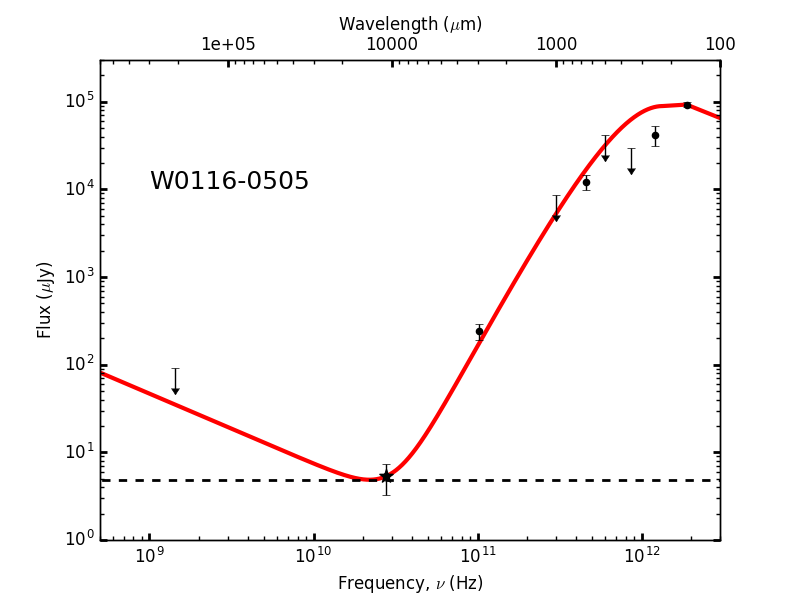}\label{W0116_SED}}\hspace{0.05cm}
\subfloat[][]{\includegraphics[trim={0.4cm 0.05cm 1.5cm 0.2cm}, clip, height=0.35\textwidth, width=0.49\textwidth] {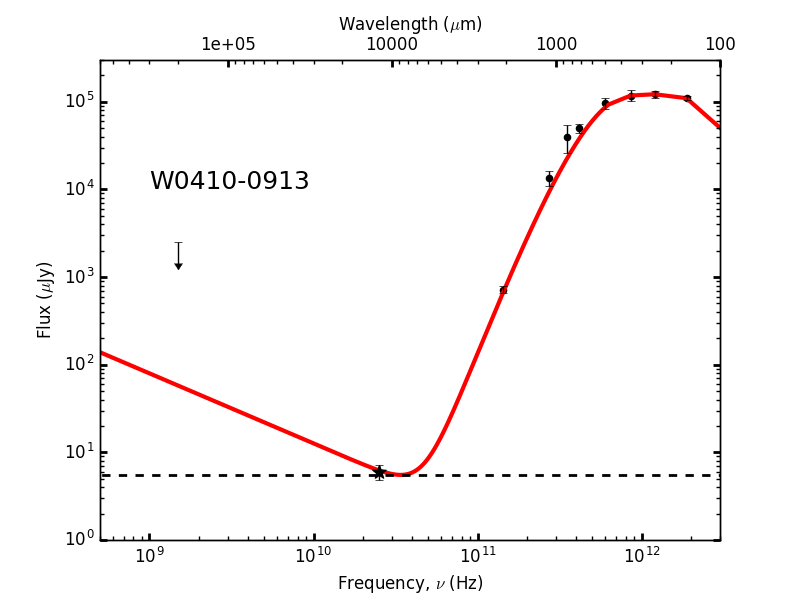}\label{W0410_SED}}\\[-6ex]
\subfloat[][]{\includegraphics[trim={0.4cm 0.05cm 1.5cm 0.2cm}, clip, height=0.35\textwidth, width=0.49\textwidth] {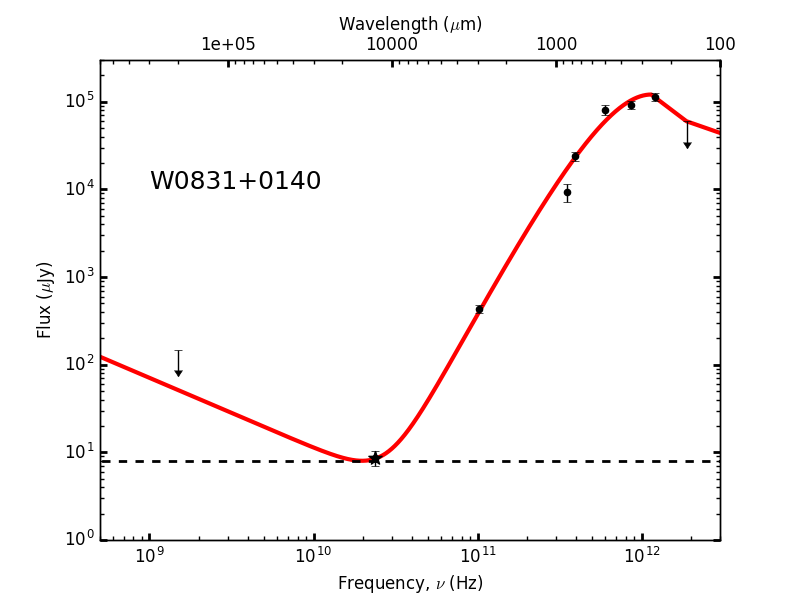}\label{W0831_SED}}\hspace{0.05cm}
\subfloat[][]{\includegraphics[trim={0.4cm 0.05cm 1.5cm 0.2cm}, clip, height=0.35\textwidth, width=0.49\textwidth] {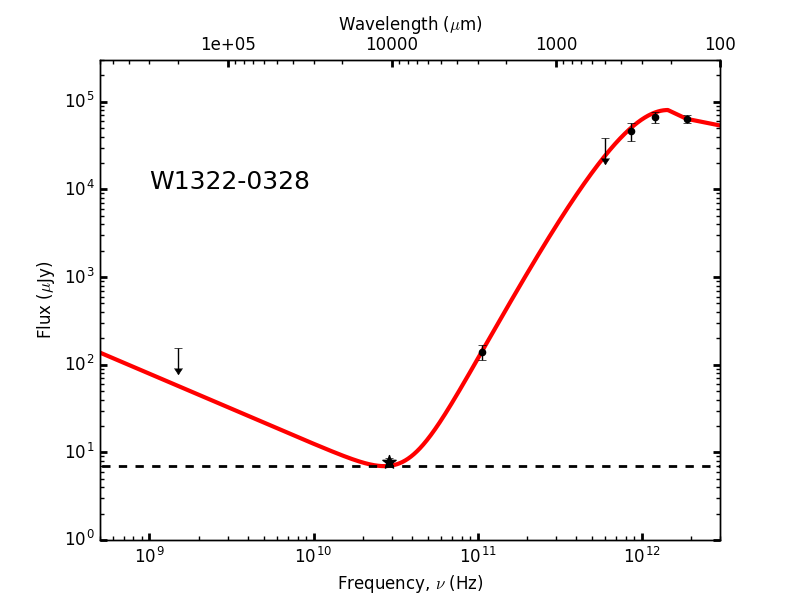}\label{W1322_SED}}\\[-3ex]
\subfloat[][]{\includegraphics[trim={0.4cm 0.05cm 1.5cm 0.2cm}, clip, height=0.35\textwidth, width=0.49\textwidth] {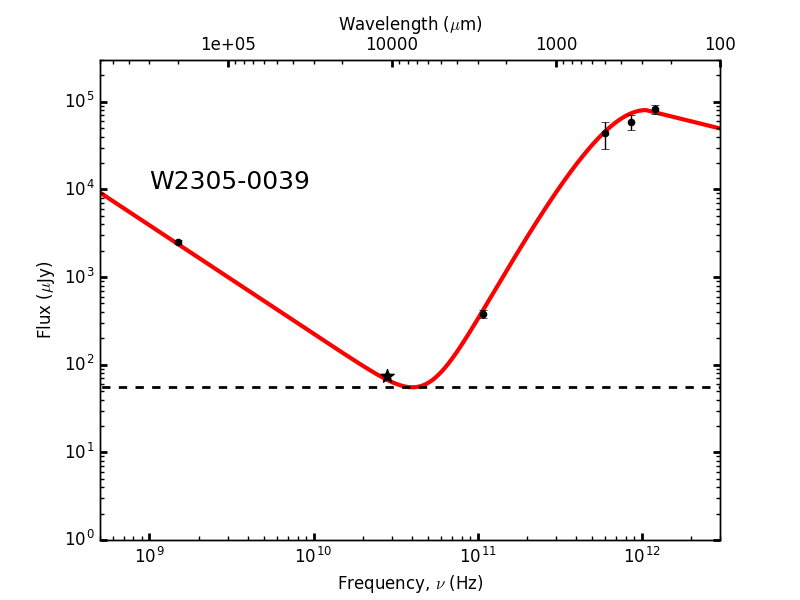}\label{W2305_SED}}\\[-5ex]
\caption{Observed frame SEDs of the five Hot DOGs in our sample. A single temperature blackbody was fitted to the $\rm160\,\mu$m to $\rm2\,$cm data, and further power-laws were added to fit the remaining data. Radio data, constrained by VLA and FIRST data, is fitted by a single power-law ($\rm S_{\nu}\propto\nu^{-0.8}$), except for W2305-0039, and the resultant curve is shown by a solid red line. A star has been used to denote the VLA data in each SED and a dashed line indicates the maximum level of free-free emission (see Table~\ref{tab:fit_params}).}
\label{SED_Images}
\end{figure*}

K- and Ka-band radio observations for high-redshift galaxies are rare, and so a comparison with related observations is difficult. Our fluxes are $\rm10-50\%$ of those obtained by \cite{MAravena16} for five lensed DSFGs selected using CO(1--0) emission within a redshift range of $z\rm=2-3$. We compute a mean continuum flux for the Hot DOGs of $\rm23.6\pm11.5\,\mu$Jy, indicated in Fig.~\ref{SED_Images}, compared with $\rm143\pm29\,\mu$Jy for the lensed DSFGs in \cite{MAravena16}. We see a similar result for when comparing two SMGs ($z\rm=2.5-3$) from \cite{APThomson12} with values of $\rm42\pm20\,\mu$Jy and $\rm57\pm25\,\mu$Jy. This suggests that the amount of cold molecular gas in these Hot DOGs is significantly lower than DSFGs and SMGs. This is surprising, given the very different nature of the cold gas in galaxies in our sample compared to typical SMGs and DSFGs at their respective redshifts.

\section{Spectral Energy Distributions}\label{SEDs}
Using archival data in addition to our VLA observations, we produce SEDs of the five Hot DOGs in our sample. We have included ALMA continuum data at $\rm\lambda_{obs}\simeq700\,\mu$m ($\rm\lambda_{obs}\simeq434\,\mu$m for W0410-0913), and $\textit{Herschel}$ PACS and SPIRE (observed frame $\rm70\,\mu$m, $\rm160\,\mu$m, $\rm250\,\mu$m, $\rm350\,\mu$m, $\rm500\,\mu$m) data from \cite{CWTsai15}. We have also added data from Warm-$\textit{Spitzer}$ IRAC1 ($\rm3.6\,\mu$m) and IRAC2 ($\rm4.5\,\mu$m) \citep{RLGriffith12}, $\textit{WISE}$ ($\rm 3.4\,\mu$m, $\rm4.6\,\mu$m, $\rm12\,\mu$m and $\rm22\,\mu$m), Bolocam ($\rm1\,mm$) \citep{JWu12} and $\rm21\,cm$ data from FIRST \citep{RLWhite97}. For W0831+0140, we include continuum data from SCUBA-2 \citep[$\rm850\,\mu$m;][]{SFJones14} and ALMA CII ($\rm\sim387\,GHz$) observations (Gonz{\'a}lez-L{\'o}pez et al. in prep). For W0410-0913, we include SHARC-II $\rm350\,\mu$m observations \citep[see][]{JWu12}. The resulting SEDs are presented in Fig.~\ref{SED_Images}.

\begin{figure}
\centering
\includegraphics[trim={0.0cm 0.4cm 0.0cm 0.0cm},clip,width=\columnwidth]{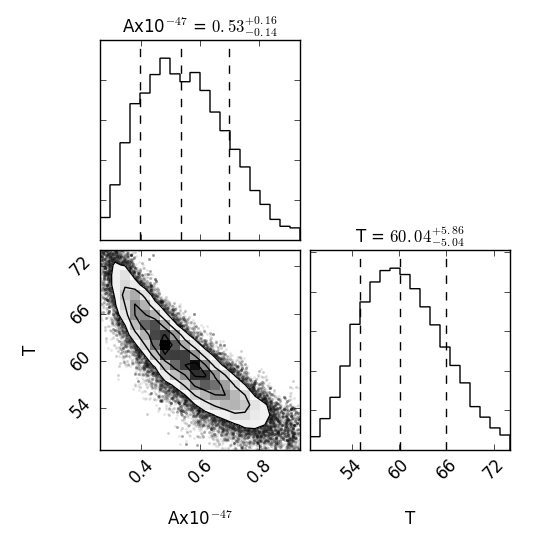}\\[-2ex]
\caption{The fitted FIR-radio SED parameters for a member of our sample (W0831+0140) using the single-temperature blackbody model. Dashed lines show the $\rm-1\sigma$, median and $\rm1\sigma$. The $A$ parameter is the normalisation and $T$ is the temperature (K).}
\label{SED_corner}
\end{figure}

We fit two separate models to the data, initially fitting a modified blackbody given by $\rm S_{\nu}\propto \nu^{3+\beta}/ (e^{\frac{h\nu}{k_{B}T_{dust}}}-1)$ between $\rm300\,GHz\lesssim\nu_{obs}\lesssim2\,THz$. This frequency range is used in all of the SEDs except W0410-0913, in which we see a different distribution of dust temperatures than the rest of the targets. Thus, we chose to fit the modified blackbody between $\rm300\,GHz\lesssim\nu_{obs}\lesssim0.7\,THz$ for this target. After the fitting is performed, the dust emission is extrapolated to $\rm\nu=115\,GHz$ and subtracted from the VLA flux, and after this a single power-law is fitted ($\rm\nu_{obs}<300\,GHz$) to estimate the synchrotron emission from these galaxies. A standard Monte-Carlo Markov Chain (MCMC) Metropolis-Hastings routine \citep[see e.g.;][]{CPRobert15} is used to fit both models to the data, and the resulting curves from the blackbody and synchrotron power-law are combined with power-laws at $\rm\nu_{obs}\gtrsim2\,THz$ to produce the final spectral fit (Fig.~\ref{SED_Images}). A single blackbody fit is necessary to accurately model the dust at increasing $\rm T_{dust}$, and for consistency with previous work, we fix $\rm\beta=1.5$ \citep{AWBlain03,JWu12} for all galaxies except W0410-0913, in which we find $\rm\beta\sim3$, due to the lower frequency peak in the SED. The resulting error in the normalisation and temperature parameters are shown in Fig.~\ref{SED_corner} and Table~\ref{tab:fit_params} for the blackbody fit. 

As described above, we use a power-law to fit the FIRST and dust-subtracted VLA continua. Given the limited constraint provided by the data, we fix the slope to a value of $\rm\alpha=-0.8$, typical for synchrotron emission \citep{JJCondon92,CLCarilli99,BCLacki10}, except for W2306-0039, for which we find a value of $\rm\alpha=-1.23\pm0.02$. Upon determining the combined synchrotron and FIR-radio fits, we also attempt to determine the maximum level of free-free emission of these galaxies, assumed to follow the form $\rm S_{\nu}\propto\nu^{-0.1}$, which could dominate between dust-dominated and synchrotron-dominated emission. The maximum permitted fluxes for the free-free emission, shown in Table~\ref{tab:fit_params}, are affected by our assumption of the value of $\alpha$ and the error in the parameters for the FIR-radio blackbody model (see Fig.~\ref{SED_corner}).

\begin{table*}
    \centering
    \caption{The fitted temperatures for the FIR-radio blackbody curve between $\rm160\,\mu$m and $\rm2\,$cm in Fig.~\ref{SED_Images} and the inferred $\rm21\,cm$ and $\rm850\,\mu$m fluxes. All values in this table are inferred from the fits in Fig.~\ref{SED_Images}. We show the maximum possible flux contribution from free-free emission ($\rm S_{ff}$). The final two columns show the $q_{\rm IR}$ values, discussed in Section~\ref{qIR} (Eqn.~\ref{equation:qIR}) and the reduced $\rm\chi^{2}$ values for the SED fits. Data marked with ($\rm^{\alpha}$) have fitted points which can be compared with observations.}
    \begin{tabular}{c c c c c c c c c}
         \hline\\[-2ex]
         ID & T (K) & $\rm S_{21\,cm}$ ($\mu$Jy) & $\rm S_{850\,\mu m}$  (mJy) & $\rm S_{ff}$ ($\mu$Jy) & $\rm L_{bol}$ ($\rm10^{13} L_{\odot}$) & $\rm L_{IR}$ ($\rm10^{13} L_{\odot}$) & $q_{\rm IR}$ & $\rm\chi^{2}$\\
         \hline\\[-2ex]
	     W0116$-$0505 & 69$\rm\substack{+4 \\ -5}$ & 36.5$\pm$9.2                   & 8.4$\pm$3.6               & $\rm<4.87$  & 13.4  & 9.13 & $\rm>2.63$ & 1.49 \\[2ex]
	     W0410$-$0913 & 32$\pm$3                   & 62.3$\pm$6.7                   & 23$\pm$3                  & $\rm<5.53$  & 17.7  & 11.2 & 2.75$\pm$0.11 & 0.75 \\[2ex]
	     W0831$+$0140 & 60$\rm\substack{+6 \\ -5}$ & 55.3$\pm$10.3                  & 17$\pm$2.7 $\rm^{\alpha}$ & $\rm<8.01$  & 18.9  & 11.2 & $\rm>2.35$ & 0.69\\ [2ex]
	     W1322$-$0328 & 62$\rm\substack{+6 \\ -4}$ & 61.7$\pm$5.5                   & 6.1$\pm$2.3               & $\rm<6.96$  & 10.7  & 7.06 & $\rm>2.30$ & 0.13\\[2ex]
	     W2305$-$0039 & 47$\pm$3                   & 2668.6$\pm$519.6$\rm^{\alpha}$ & 14$\pm$1.6                & $\rm<55.28$ & 14.4  & 7.57 & 0.85$\pm$0.22 & 0.37\\
	     \hline
    \end{tabular}
    \label{tab:fit_params}
\end{table*}

From the SEDs of the five Hot DOGs in our sample, we find that most $\rm20-30\,$GHz VLA continuum fluxes are consistent with synchrotron and free-free emission ($\rm S_{\nu}\propto\nu^{-0.1}$), with only a small potential contribution from dust in a few cases. Without additional radio fluxes, we are unable to determine which feature is dominant at rest-frame $\rm115\,GHz$ for these galaxies. W2305-0039 is the exception. W2305-0039 has a significant FIRST detection, likely caused by additional synchrotron emission, possibly dominating thermal dust in the VLA observations. As shown by the fit in Fig.~\ref{SED_Images}, it is likely that the rest-frame $\rm115.3\,GHz$ free-free emission in W2305-0039 is dominated by synchrotron emission.

\begin{figure}
\centering
\includegraphics[trim={0.0cm 0.0cm 0.0cm 0.0cm},clip,width=\columnwidth]{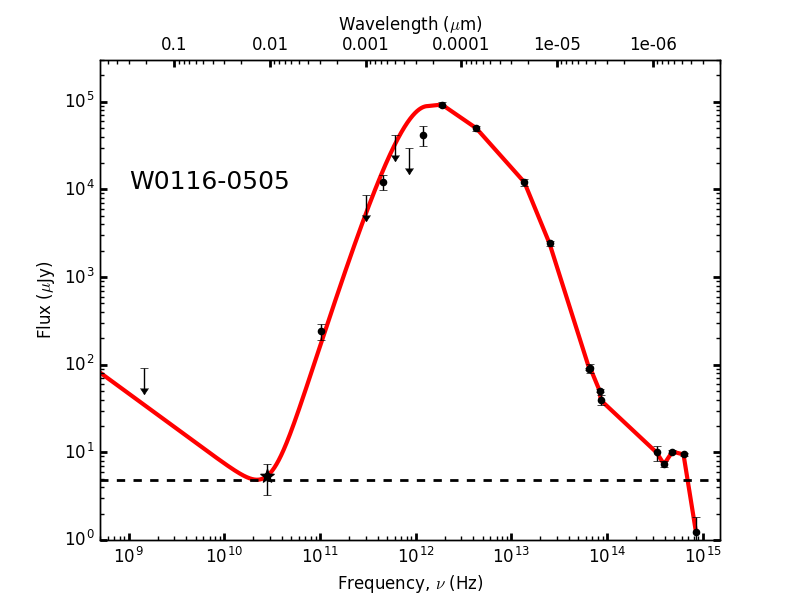}\\[-2ex]
\caption{The full SED for an example galaxy (W0116-0505). The values from the fits were used to determine the K-corrected FIR luminosity, shown in Section~\ref{qIR}.}
\label{W0116_SED_Full}
\end{figure}

We list the $\rm21\,cm$ fluxes from the synchrotron fits at $\rm\nu_{obs}\lesssim300\,GHz$ with $\rm1\sigma$ errors in Table~\ref{tab:fit_params}, using a fixed spectral slope of $\rm\alpha=-0.8$. The results suggest that the expected fluxes are close to the specific RMS of the FIRST images at the location of our targets, except for W0410-0913, which only appears in the area of the NRAO VLA Sky Survey \citep[NVSS;][]{JJCondon98}, These values are significantly lower than the reported depth of the FIRST catalogue ($\rm\sim1\,mJy$) by a factor of 8--30, and are likely to be more precise upper limits for all galaxies in our sample, except W0410-0913. Recent findings by \cite{ALaor19} suggest that galaxies with a higher Eddington ratio may have a steeper spectral index, and given that \cite{CWTsai15} found that these Hot DOGs should have an Eddington ratio $\rm\sim1$, it is likely that the $\rm21\,cm$ flux estimate for W0410-0913, shown in Table~\ref{tab:fit_params}, is an underestimate. This is further evidenced by the only target detected in FIRST, W2305-0039, which has a fitted spectral index of $\rm\alpha=-1.23\pm0.02$. Deeper cm-wavelength observations could be used to understand the nature of the synchrotron emission in these galaxies and also further constrain maximum fluxes from free-free emission and $\rm21\,cm$ emission.

Using the fits for the full spectrum (e.g. Fig.~\ref{W0116_SED_Full}), we estimate the bolometric and IR luminosities to compare with previous work (see Table~\ref{tab:fit_params}). For the IR luminosity, measured between $\rm8\,\mu$m and $\rm1\,$mm, we see no significant deviation in the luminosity estimate compared with \cite{CWTsai15} for all galaxies in our sample. The small increase is likely due to the data at lower-frequencies included in this work and the small deviation in the blackbody fits between these two works.

Comparing the dust temperatures between $\rm160\,\mu m-20\,cm$ from our modified blackbody fits to \cite{JWu12} for W0116-0505 and W0410-0913, we find lower values. \cite{JWu12} used a similar blackbody fitting method for $\rm8\,\mu$m to $\rm1000\,\mu$m data, and found dust temperatures of $\rm123\pm8\,$K and $\rm82\pm5\,$K for W0116-0505 and W0410-0913 respectively. These are 1--3 times the values shown in Table~\ref{tab:fit_params}. While our results differ significantly from \cite{JWu12}, we fit the coldest dust using previously unavailable data from ALMA and the VLA. Furthermore, our results focus on the temperature of the coldest dust on the Raleigh-Jeans tail, and it is unlikely that our results significantly underestimate the temperature of the coldest dust in these galaxies.

\begin{figure*}
\captionsetup[subfigure]{labelformat=empty}
\subfloat[][]{\includegraphics[trim={0.0cm 0.0cm 0.0cm 0.0cm}, clip, width=0.4\textwidth]{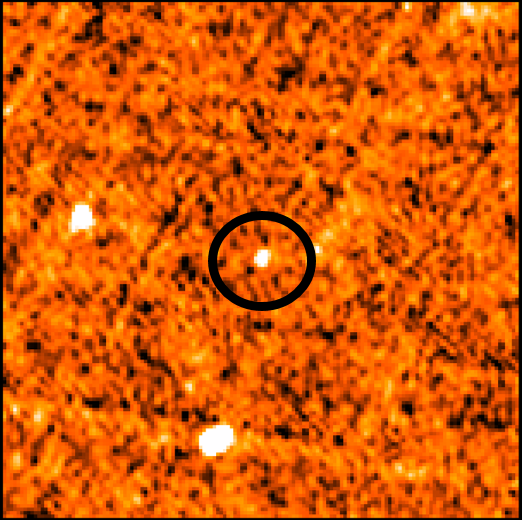}\label{W2305_FIRST}}\hspace{2ex}
\subfloat[][]{\includegraphics[trim={22.8cm 5.5cm 23.6cm 11.3cm}, clip, width=0.4\textwidth]{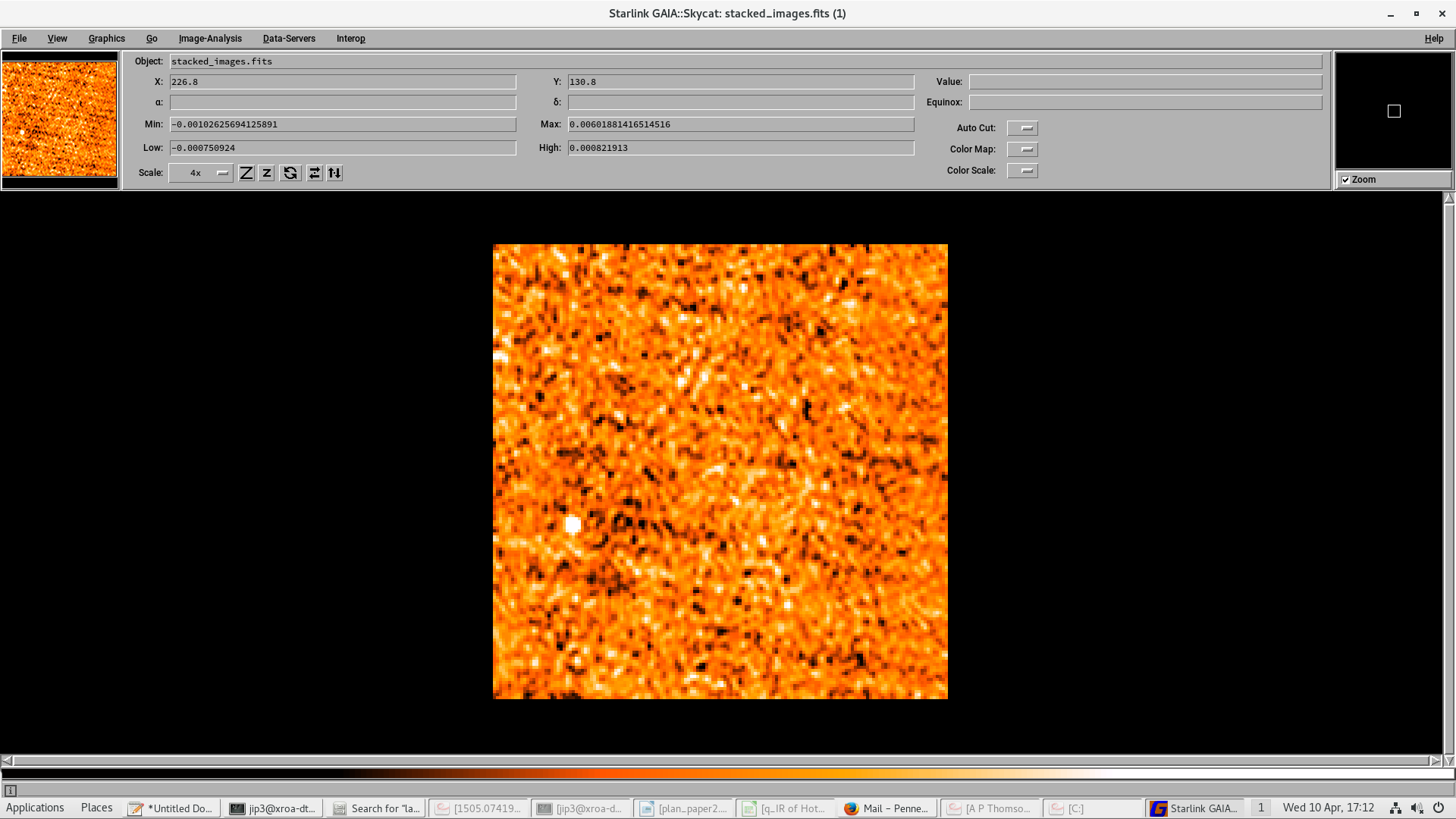}\label{Stacked_FIRST}}\\[-3ex]
\caption{Left: Image of the W2305-0039 field using FIRST image cutout service. The Hot DOG in the field has been ringed. Right: Stacked FIRST images of W0116-0505, W0126-0529, W0831+0140 and W1322-0328, showing that no net flux is visible at 1.4GHz. W0410-0913 has not been included in this process as it is outside the FIRST survey coverage. Both images constitute $\rm4.5\,'\times4.5\,'$ fields.}
\label{FIRST_Images}
\end{figure*}

Fig.~\ref{SED_Images} shows that W2305-0039 is the only galaxy of our sample with significant $\rm1.4\,$GHz radio emission. We fit a spectral index of $\rm\alpha=-1.23\pm0.02$ by combining FIRST fluxes with our VLA observations, significantly steeper than most values of $\alpha$ \citep{JJCondon92}. This spectral index is significantly greater than that determined via a stacking analysis of $z\rm\sim2$ galaxies in \cite{BMagnelli15}, and using the definition for radio-loudness in \cite{DStern00}, W2305-0039 is radio-loud compared to the other Hot DOGs in our sample. This is not surprising, given that $\rm10-15\%$ of galaxies are expected to be radio-loud \citep{RMKratzer15}, and could suggest that there is a subset of radio-loud Hot DOGs.

The maximum value of the free-free emission is $\rm\lesssim3\,\mu$Jy for the Hot DOGs in our sample, and significantly greater for W2305-0039 due to the lower value of $\alpha$ used, though its value is likely to be much lower given the dominance of its synchrotron emission. The maximum values are lower than the values calculated in \cite{APThomson12}, which found, for two SMGs, values of $\rm12\,\mu$Jy and $\rm24\,\mu$Jy. This implies that the Hot DOGs have lower free-free emission than typical SMGs and suggests that these galaxies could have lower star-formation rates than SMGs at lower redshifts. However, the limited number of sources precludes a statistical assessment of whether this is characteristic of the entire Hot DOG population, and if so, whether this is due to the greater activity of the AGN or the surrounding environment.

\section{Discussion}\label{Discussion}
\subsection{$\rm30\,$GHz Emission from Hot DOGs}\label{Radio_Emission_Hot_DOGs}
Our VLA data for five Hot DOGs are near the expected continuum flux from a dust-emission power-law ($\rm S \propto\nu^{2+\beta}$), and we find that dust emission could provide a mean $\rm\sim20\%$ of the VLA flux. At $\rm1.4\,$GHz, we predict a distinct lack of emission in 4/5 galaxies in our sample, as shown in Fig.~\ref{SED_Images}. To understand whether there is any emission just beneath the noise level of FIRST, we stack the FIRST images for W0116-0505, W0126-0529, W0831+0140 and W1322-0328. The Root-Mean-Square (RMS) values for the individual images are a third of the $\rm1.4\,GHz$ flux for W2305-0039, however we see no stacked detection (Fig.~\ref{FIRST_Images}), suggesting that the $\rm1.4\,$GHz emission of these galaxies is substantially lower than the detection limit of FIRST. It is unlikely that the dust emission from these Hot DOGs is suppressing the synchrotron processes at $\rm1.4\,$GHz, and it is more likely that these Hot DOGs are relatively radio-quiet compared to W2305-0039.

\subsection{FIR-Radio Correlation}\label{qIR}
The FIR-radio correlation has been widely used to identify star formation properties of galaxies at high redshift \citep[see][for a review]{BCLacki10Hz}. Star formation drives the correlation, in which dust-obscured stellar emission produces flux in the FIR, while SNe from the same stars with masses $\rm M_{\odot}\gtrsim8M_{\odot}$ produce synchrotron emission at $\rm\sim1.4\,$GHz. The $q_{\rm IR}$ value is used to compare the difference in the emission, given by:
\begin{equation}\label{equation:qIR}
    \rm q_{IR}=\log_{10}\left(\frac{L_{FIR}}{3.75\times10^{12}\times S_{\rm 1.4\,GHz}}\right) ,
\end{equation}

\noindent where $\rm L_{FIR}$ is the rest-frame FIR luminosity in $\rm W\,m^{-2}$, and $\rm S_{1.4\,GHz}$ is the flux at rest-frame $\rm1.4\,$GHz in $\rm W\,m^{-2}\,Hz^{-1}$. The correlation is well documented locally \citep{PCKruit71,GHelou85,JJCondon92,MSYun01}, and extending to higher redshifts \citep{BCLacki10,RJIvison10}, with \cite{APThomson14} finding $q_{\rm IR}=2.56\pm0.05$ using ALMA $\rm870\,\mu$m observations for $\rm\sim50$ $z\rm\gtrsim2$ SMGs. \cite{RJIvison10} found a similar $q_{\rm IR}=2.40\pm0.24$ for $\textit{Herschel}$ $\rm250\,\mu$m-selected galaxies at $z>\rm1$, and \cite{BJBoyle07} discovered that the correlation was constant down to $\mu$Jy fluxes for radio-quiet quasars.

To estimate the value of the $q_{\rm IR}$ for the FIRST undetected Hot DOGs in our sample, we assume $S\propto\nu^{\alpha}$ where $\rm\alpha=-0.8$ for synchrotron emission to extrapolate the expected value at $\rm1.4\,$GHz from the VLA detections. This extrapolation is necessary due to their non-detections (for 4/5 galaxies) in FIRST. As previously discussed, \cite{ALaor19} suggest that galaxies with a higher Eddington luminosity may have steeper spectral indices, and could suggest that we have underestimated the $\rm1.4\,$GHz fluxes of our sample. However, without additional observations, we cannot determine the true value of the rest-frame $\rm1.4\,$GHz emission. We see no significant deviation in the $q_{\rm IR}$ values for these galaxies using the $\rm L_{FIR}$ values from \cite{CWTsai15}, which is expected given that our $\rm L_{FIR}$ values deviate by only $\rm\sim15\%$ at most. 

We compute the $q_{\rm IR}$ values for the Hot DOGs in our sample using the SED fits of the FIR within the wavelength range $\rm8-1000\,\mu$m \citep{RJIvison10}, as shown in Table~\ref{tab:fit_params}. All of the galaxies undetected in FIRST/NVSS possess $q_{\rm IR}$ estimates within the upper $\rm2\sigma$ limits from \cite{RJIvison10}, $q_{\rm IR}\rm=2.40\pm0.24$, suggesting that these galaxies may be radio-quiet. W0410-0913, undetected in NVSS, has a $q_{\rm IR}$ value estimated using the assumption that the radio spectral index is $\rm\alpha=-0.8$. However, it should be noted that, given the more relaxed constraints in NVSS, the $q_{\rm IR}$ value could be significantly affected by this assumption, and may possess a $q_{\rm IR}$ value similar to W2305-0039. Using our assumption, we see that W0410-0913 has a $q_{\rm IR}$ value above the $\rm2\sigma$ limits from \cite{RJIvison10}, suggesting that this Hot DOG is radio-quiet with respect to the FIR emission. The lack of radio-emission from these undetected galaxies is unusual, given that the obscured emission in the FIR from the presence of an AGN should be expected to also radiatively boost the radio emission. The higher $q_{\rm IR}$ values from W0116-0505 and W0410-0913 are likely due to dust significantly boosting the FIR emission, while the radio emission is unaffected. The $q_{\rm IR}$ value for W2305-0039 ($q_{\rm IR}=0.85\pm0.22$) is at least an order of magnitude smaller than the mean $z\rm=2$ value from \cite{RJIvison10}, suggesting that W2305-0039 is more radio-loud than expected for SMGs, and thus that the galaxies in our sample are all different from the population observed in previous work on the high-redshift FIR-radio correlation.

\cite{DRGSchleicher13} and \cite{JSchober16} suggest a break down of the FIR-radio correlation at high redshifts due to more effective electron cooling by Bremsstrahlung emission in the cosmic microwave background. This is unlikely to be the cause of the increased $q_{\rm IR}$ values in the NVSS/FIRST undetected galaxies in our sample, given that the radiation field of the galaxies will dominate over the cosmic microwave background, and this effect should only dominate at $z\rm\gtrsim5$. The higher $q_{\rm IR}$ values for the Hot DOGs in our sample are likely due to the increased FIR emission from the dust-obscured AGN shown in previous work \citep{RJAssef15,TDSantos16}, such that the AGN boosts the FIR-emission from young stars compared to normal galaxies.

If the central AGN is radio-quiet, the observed $q_{\rm IR}$ values from these galaxies would be dramatically increased. This is unexpected and it is likely that the dust has boosted the FIR emission above what would be expected, while the $\rm1.4\,$GHz emission is similar to that of a normal galaxy at this redshift. A lack of synchrotron emission could also suggest an abundance of young stars that have not undergone a SNe event within these galaxies, and could suggest that these galaxies are still forming stars, which could explain the lower rest-frame $\rm115.3\,$GHz continuum fluxes observed in Table~\ref{tab:Luminosity2}. \cite{BMagnelli15} and \cite{JDelhaize17} suggest a weak redshift dependence in the FIR-radio correlation of the form $q_{\rm IR}$($z$)$\rm=q_{\rm IR}$(1+$z$)$\rm^{-\beta}$, where $\beta\sim0.15$. Using the values from \cite{RJIvison10}, we expect to see $q_{\rm IR}(z)\sim2\pm0.15$ for the Hot DOGs in our sample, implying significantly greater radio fluxes than the FIRST limits suggest. These results were derived for star-forming galaxies, and so the FIR may be significantly enhanced by AGN emission in our sample, increasing the expected FIR emission beyond the radio emission produced from SNe. This agrees with previous results from SED fits of Hot DOGs \citep{SFJones14, RJAssef15}, and further suggests that the AGN in these galaxies are radio-quiet.

\begin{figure}
\centering
\includegraphics[trim={1.0cm 0.0cm 1.4cm 1.0cm}, clip, width=\columnwidth]{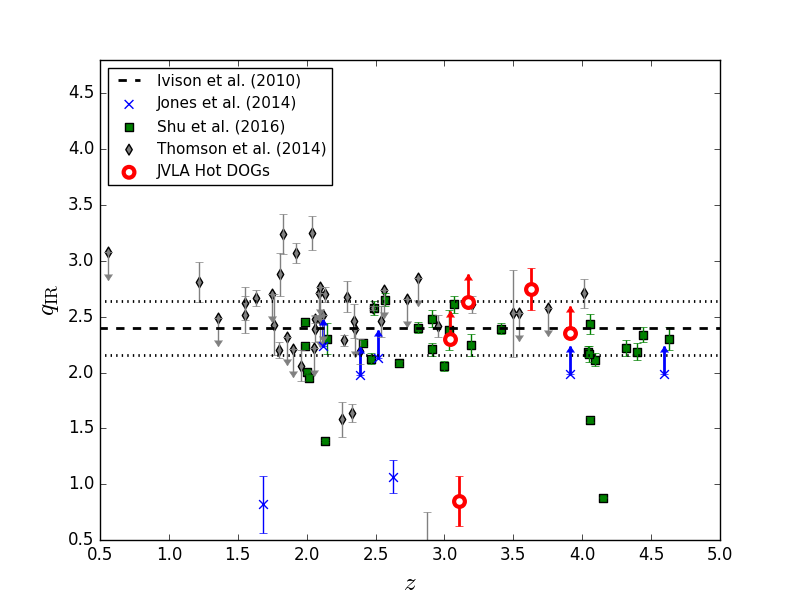} \\[-2ex]
\caption{Comparison of $q_{\rm IR}$ values derived using rest-frame $\rm L_{8-1000\,\mu m}$ and $\rm S_{1.4\,GHz}$ for the Hot DOGs in our sample (red open circles) against previously observed Hot DOGs \citep{SFJones14} (blue crosses). $\rm L_{8-1000\,\mu m}$ was derived using the fits from the SEDs in Fig.~\ref{SED_Images}. $\rm S_{1.4\,GHz}$ values were K-corrected, using a spectral index of $\rm\alpha=-0.8$. Comparison samples are shown by green squares and grey diamonds for star-forming DSFGs \citep{XWShu16} and $\rm870\,\mu$m selected SMGs \citep{APThomson14} respectively. A black dashed line and black dotted lines illustrate the median value and the $\rm\pm1.5\sigma$ values for $\rm250\,\mu$m selected galaxies in the GOODS-N field \citep{RJIvison10}.}
\label{q_IR_Image}
\end{figure}

To understand whether the $q_{\rm IR}$ values for our sample are typical for Hot DOGs or other classes of object, we compare our values with both high and low redshift galaxies, comparing $q_{\rm IR}$ values determined using the FIR luminosity between $\rm8-1000\,\mu$m. We compare with 10 Hot DOGs from \cite{SFJones14}, which used $\rm850\,\mu$m SCUBA-2 data to search for companion SMGs around Hot DOGs. W0831+0140 is common to both works, and we expect that the Hot DOGs in \cite{SFJones14} have similar radio properties to the Hot DOGs in this work. To understand whether our $q$-values are typical of other dusty galaxies at this redshift, we compare with star-forming DSFGs from \cite{XWShu16} selected from the GOODS-N field, which have redshifts $z\rm=2-4$. We also compare with 52 ALMA $\rm870\,\mu$m selected SMGs in the Extended Chandra Deep Field South with a redshift range of $z\rm_{phot}\sim0.4-4.7$ (median $z\rm=2.125$) from \cite{APThomson14}, which were observed using the VLA at $\rm1.4\,$GHz at $\rm>3\sigma$ significance. The results are shown in Fig.~\ref{q_IR_Image}, which illustrates the distribution of the $q_{\rm IR}$ values as a function of redshift.

Hot DOGs undetected in the FIRST survey from \cite{SFJones14} have $q_{\rm IR}$ values consistent with the Hot DOGs undetected in the FIRST survey from this work, suggesting that the non-detection of $\rm1.4\,$GHz radio emission is typical for this class of dust-obscured galaxies. We find two Hot DOGs with detections in \cite{SFJones14}, which have similar $q_{\rm IR}$ values to W2305-0039, suggesting a possible sub-class of radio-loud Hot DOGs. It should be noted that these Hot DOGs have different selection criteria than the radio-$\textit{WISE}$ selected galaxies in \cite{CJLonsdale15}, as shown by the selection comparison in \cite{JIPenney19}. There are Hot DOGs in both works without detections in FIRST, with lower limits for the q-values that agree with the median values for \cite{RJIvison10} and \cite{APThomson14}, suggesting that these galaxies could agree with the FIR-radio correlation. However, two Hot DOGs from this work (W0116-0505 and W0410-0913) have lower limits for their $q_{\rm IR}$ values close to the upper limit in \cite{APThomson14}, suggesting that the lower limits for the Hot DOGs could severely underestimate their true $q_{\rm IR}$ value. We see agreement between the three comparison works, showing that the distribution of $q_{\rm IR}$ values is observed in other galaxies with a wide range of selections and redshifts, but is generally not observed in Hot DOGs from this work and \cite{SFJones14}, and that Hot DOGs have a wider spread of $q_{\rm IR}$ values than typical galaxies. It should be noted that the Hot DOGs in this work, except W0831+0140, differ from the Hot DOGs in \cite{SFJones14} due to the luminosity constraints in \cite{CWTsai15}, but are expected to be the same class of galaxy. These results suggest that Hot DOGs are radio-quiet and the increased FIR luminosity has likely boosted the expected $q_{\rm IR}$ values within the upper $\rm2\sigma$ values.

\section{Conclusions}\label{Conclusions}
The results from the VLA observations of five WISE-selected, dusty, luminous obscured galaxies are:
\begin{itemize}
    \item We formulate SEDs of the five Hot DOGs in our sample from archival data in conjunction with ALMA and VLA observations and a single modified blackbody and power-law models. We find that synchrotron/free-free emission dominates at the rest-frame $\rm115.3\,$GHz for four of the galaxies in our sample, which show little or no dust contribution, and W2305-0039 contains additional synchrotron emission, increasing the rest-frame $\rm115.3\,GHz$ fluxes above a single temperature blackbody model.
    \item We determine the CO(1--0) luminosity for the five Hot DOGs in our sample and compare with $z\rm\sim2$ SMGs from \cite{RJIvison11}. We find that the Hot DOGs have significantly lower values for the CO(1--0) line luminosity, and may not possess significant cold molecular gas reservoirs compared with lower redshift, star-forming galaxies.
    \item We estimate the maximum free-free emission in the galaxies in our sample and compare to two $z\rm\sim2$ SMGs from \cite{APThomson12}. We find that the Hot DOGs have typically lower fluxes for free-free emission and may not possess as significant star formation activity as SMGs at lower redshifts.
    \item We determine that the $q_{\rm IR}$ values for the five Hot DOGs in our sample using SED fits, and find the galaxies undetected in FIRST have higher values compared with other galaxies at these redshifts. This suggests that the FIR has been boosted by the dust emission, significantly increasing $q_{\rm IR}$. Thus, Hot DOGs are generally radio-quiet, and are dominated by strong FIR emission.
    \item We determine that W2305-0039 has an additional $\rm115.3\,$GHz CO(1--0) radio companion. W2305-0039 likely has compact radio emission, rather than jets extending from the host galaxy, and is more radio-loud than the other Hot DOGs in our sample.
    \item We find redshifts in agreement with previous work by \cite{CWTsai15} for 4 Hot DOGs. For W0410-0913, we find a redshift outside the predicted error margins from previous work, but in agreement with new findings using ALMA. We find no CO(1--0) emission line in the spectra of W0126-0529 at the frequency corresponding to a previously known redshift, which was later updated to $z\rm=0.8301$ \citep{HJun20}.
\end{itemize}

\section*{Acknowledgements}
The authors wish to thank the staff in the Department of Physics and Astronomy at the Leicester University for their support. Jordan Penney is supported by the Science and Technologies Facilities Council (STFC) studentship. M. Kim was supported by the National Research Foundation of Korea (NRF), grant funded by the Korean government (MSIP) (No. 2017R1C1B2002879). R. J. Assef was supported by the FONDECYT grant (NO. 1191124). T.D-S. acknowledges support from the CASSACA and CONICYT fund CAS-CONICYT Call 2018. This research was carried out in part at the Jet Propulsion Laboratory, California Institute of Technology, under a contract with NASA. H.D.J. was supported by the Basic Science Research Program through the National Research Foundation of Korea (NRF) funded by the Ministry of Education (NRF- 2017R1A6A3A04005158). C.-W. Tsai was supported by a grant from the NSFC (No. 11973051). The National Radio Astronomy Observatory is a facility of the National Science Foundation operated under cooperative agreement by Associated Universities, Inc. This publication makes use of data products from the $\textit{Wide-field Infrared Survey Explorer}$, which is a joint project of the University of California, Los-Angeles, and the Jet Propulsion Laboratory/California Institute of Technology, funded by the National Aeronautics and Space Administration. This paper makes use of ALMA data ADS/JAO.ALMA$\#$2017.1.00358.S. ALMA is a partnership of ESO (representing its member states), NSF (USA) and NINS (Japan), together with NRC (Canada), NSC and ASIAA (Taiwan), and KASI (Republic of Korea), in cooperation with the Republic of Chile. The Joint ALMA Observatory is operated by ESO, AUI/NRAO and NAOJ.




\bibliographystyle{mnras}
\bibliography{Radio_Hot_DOGs_CO(1_0)} 




\appendix
\section{Appendix}\label{Appendix}
Here, we discuss the results for observations of W0126-0529, previously expected to be a Hot DOG at $z\rm=2.937$, later determined to be at $z\rm=0.8301$ \citep{HJun20} after the initial VLA observations were made. Given this lower redshift, W0126-0529 is not part of the subset of Hot DOG discussed in \cite{PRMEisenhardt12} and has therefore been omitted from the main results of this paper. In Table~\ref{tab:W0126_params}, we provide measurements for W0126-0529 given the new redshift information.

\begin{figure}
    \centering
    \includegraphics[trim={0.0cm 0.0cm 7.4cm 0.0cm},clip,width=0.9\columnwidth]{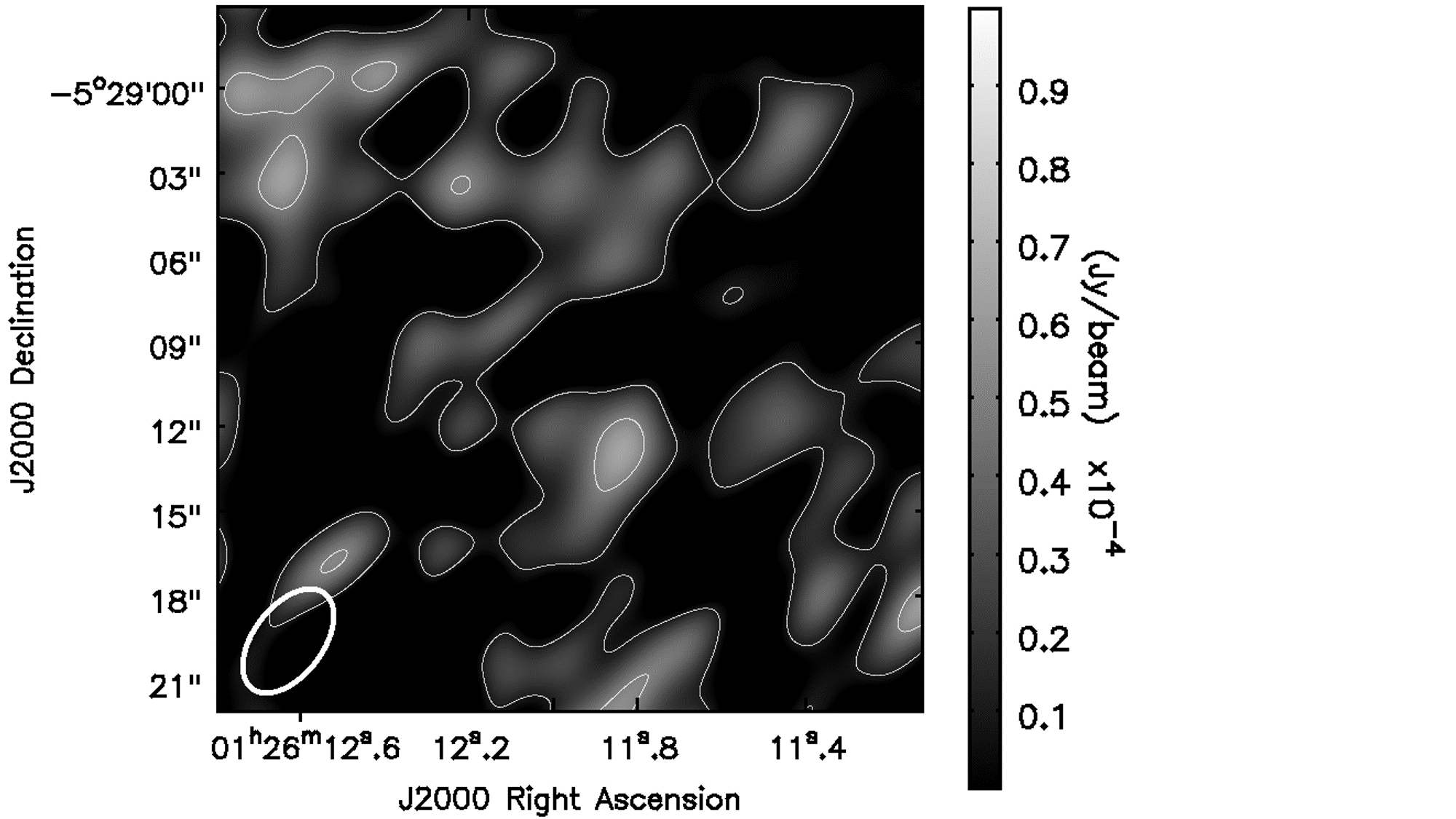}\\[2ex]
    \includegraphics[trim={0.0cm 0.0cm 6.4cm 0.0cm},clip,width=0.9\columnwidth]{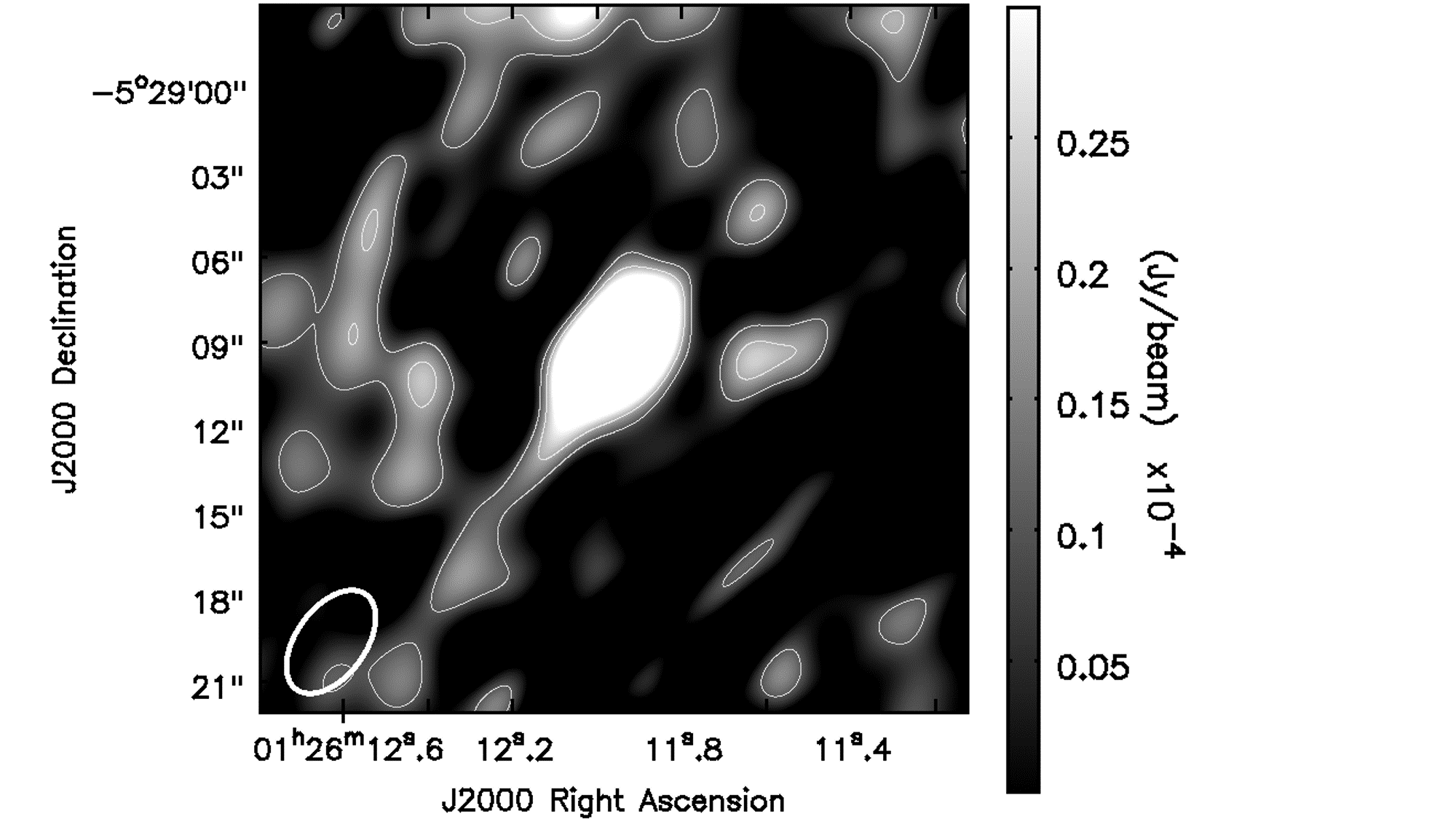}\\
    \caption{Emission line (top) and continuum image (bottom) for W0126-0529. Contours show the $\rm0.1\,\mu$Jy, $\rm0.5\,\mu$Jy, $\rm1\,\mu$Jy levels for the emission line image and the $\rm0.1\,\mu$Jy, $\rm0.2\,\mu$Jy and $\rm0.3\,\mu$Jy levels for the continuum image. An ellipse in the bottom left-hand corner of both images illustrates the synthesized beam for the image.}
    \label{W0126_maps}
\end{figure}

We use the same calibration pipeline noted in Section~\ref{Observations}, and use the same \textsc{tclean} parameters in CASA to produce the spectral and continuum images, shown in Fig.~\ref{W0126_maps}. We use the $spw$ corresponding to the redshift from \cite{CWTsai15}. As expected, we find no CO(1--0) emission line at this frequency ($\rm\nu_{obs}\sim29.29\,$GHz) as shown in Fig.~\ref{W0126_maps}, and there are no other lines within the observed frequency range for $z\rm=0.8301$, such that we cannot confirm the redshift identification of \cite{HJun20}. In Fig.~\ref{W0126_maps} we observe a potential companion source, similar to W2305-0039, $\rm\sim5\,''$, below the expected position, however we find no emission spectra at this position consistent with the redshift in \cite{CWTsai15} and this is likely to be noise. The continuum image for W0126-0529 shows a single, bright object in the centre of the field, as expected, with a flux measurement of $\rm128\pm14.6\,\mu$Jy, brighter than any of the continuum emission for the five Hot DOGs in this paper. The significantly brighter continuum emission for this object is likely due to the lower redshift, placing the VLA observations further along the Raleigh-Jean's tail, closer to the peak in the FIR emission. We model the expected spectrum for this galaxy below.

\begin{figure}
\centering
    \includegraphics[trim={0.4cm 0.0cm 1.5cm 0.0cm}, clip, width=\columnwidth]{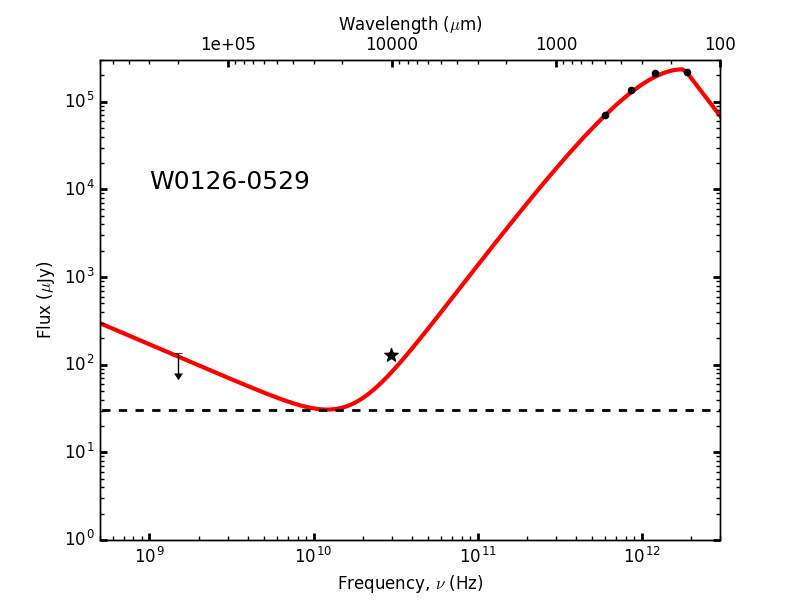}\\[-1ex]
    \caption{Observed frame SED for W0126-0529, using the same single temperature, modified blackbody fit to the $\rm160\,\mu$m to $\rm2\,cm$ data with additional power-laws to fit for the remaining data. Radio data from FIRST and the VLA is fitted using a power-law ($\rm S_{\nu}\propto\nu^{-0.8}$). The resultant curve is shown by a solid red line and a star has been used to denote the VLA data point. The dashed line indicates the maximum free-free emission possible.}
    \label{W0126_SED}
\end{figure}

Using the same method outlined in Section~\ref{SEDs}, we use archive data for W0126-0529 to construct an SED, shown in Fig.~\ref{W0126_SED} to estimate the properties of the object, listed in Table~\ref{tab:W0126_params}. Unlike the Hot DOGs in Section~\ref{SEDs}, we used a lower $\beta$ value of $\rm\beta=0.5$, which appeared to better fit the slope than the previously used value of $\rm\beta=1.5$ for the Hot DOGs. Understandably, this suggests that there is significantly less dust obscuration for this object than for the Hot DOGs in our sample, and could suggest that it is a normal-type galaxy. From the SED, we find a maximum free-free emission flux of $\rm31.3\,\mu$Jy, significantly greater than all of the Hot DOGs in our sample, except for W2305-0039, suggesting that W0126-0529 could possess large quantities of cold molecular gas. Additional observations at lower frequencies than these VLA observations are required to confirm this.

From the SED model of this galaxy, we are able to estimate the $q_{\rm IR}$ value, and find a value of $q_{\rm IR}=2.97\pm0.77$, suggesting that this galaxy is more radio-quiet than expected from previous work by \cite{RJIvison10}. This is unusual, given the large quantities of cold molecular gas residing within this object, and could suggest that this object does not have significant AGN activity, as suggested in \cite{HJun20}.

\begin{table}
    \centering
    \caption{Observed and computed properties for W0126-0529, previously believed to be a HyLIRG Hot DOG during the time of the VLA observations (see Section~\ref{Observations}). $\rm S_{1\,cm}$ denotes the continuum flux for the galaxy using the same flux measurement method in Section~\ref{Observations}, using all available $spw$s. Computed properties using the SED in Fig.~\ref{W0126_SED} below are also included.}
    \begin{tabular}{c c}
        \hline \\[-2ex]
         & W0126-0529 \\
         \hline \\[-2ex]
        RA (J2000) & 01:26:11.95 \\[2ex]
        DEC (J2000) & -05:29:09.1 \\[2ex]
        $z$ & 0.8301 \\[2ex]
        Major (arcsec) & 4.17 \\[2ex]
        Minor (arcsec) & 2.44 \\[2ex]
        PA (deg) & -36.48 \\[2ex]
        $\rm S_{1\,cm}$ ($\mu$Jy) & 128.8 $\pm$ 14.6 \\[2ex]
        $\rm \nu_{VLA}$ (GHz) & 28.7--29.7 \\[2ex]
        $\rm T_{160\,\mu m - 2\,cm}$ (K) & 20.4$\rm\substack{+0.5 \\ -0.3}$ \\[2ex]
        $\rm S_{21\,cm}$ ($\mu$Jy) & 133.6$\pm$25.3 \\[2ex]
        $\rm S_{850\,\mu m}$ (mJy) & 24$\pm$1 \\[2ex]
        $\rm S_{ff}$ ($\mu$Jy) & <30.8 \\[2ex]
        $\rm L_{bol}$ ($\rm10^{13}\,L_{\odot}$) & 0.69 \\[2ex]
        $\rm L_{IR}$ ($\rm10^{13}\,L_{\odot}$) & 0.69 \\[2ex]
        $q_{\rm IR}$ & $\rm>3.00$ \\[2ex]
        $\rm \chi^{2}$ & 1.82 \\[2ex]
        \hline
    \end{tabular}
    \label{tab:W0126_params}
\end{table}

{}


\bsp	
\label{lastpage}
\end{document}